\documentclass[preprint,3p,twocolumn,linenumbers]{elsarticle}

\usepackage[colorlinks]{hyperref}
\usepackage{tabularx}
\usepackage{graphicx}
\usepackage{amssymb,amsmath,bm,tensor,braket}
\usepackage{xcolor}
\usepackage[varg]{txfonts}
\usepackage{enumerate}
\usepackage{mathtools}
\usepackage[utf8]{inputenc}
\usepackage[normalem]{ulem}
\usepackage{lineno,hyperref}
\usepackage{cuted}
\modulolinenumbers[5]

\def\sqr#1#2{{\vcenter{\vbox{\hrule height.#2pt
         \hbox{\vrule width.#2pt height#1pt \kern#1pt
         \vrule width.#2pt}
         \hrule height.#2pt}}}}

\newcommand{\beq}{\begin{equation}}
\newcommand{\eeq}{\end{equation}}
\newcommand{\bea}{\begin{eqnarray}}
\newcommand{\eea}{\end{eqnarray}}
\newcommand{\rf}[1]{(\ref{#1})}

\newcommand{\bM}{\begin{pmatrix}}
\newcommand{\eM}{\end{pmatrix}}

\begin{document}

\begin{frontmatter}

\title{Bumblebee cosmology: The FLRW solution and the CMB temperature anisotropy}


\author[1]{Rui Xu}
\author[1]{Dandan Xu\corref{cor1}}\ead{dandanxu@tsinghua.edu.cn}
\author[2]{Lars Andersson}
\author[3,4,5]{Pau Amaro Seoane}
\author[6,7]{Lijing Shao}

\cortext[cor1]{Corresponding author}
\address[1]{Department of Astronomy, Tsinghua University, Beijing 100084, China}
\address[2]{Beijing Institute of Mathematical Sciences and Applications, Beijing 101408, China}
\address[3]{Institute for Multidisciplinary Mathematics, Universitat Politècnica de Val\`{e}ncia, Val\`{e}ncia 46022, Spain}
\address[4]{Max-Planck-Institute for Extraterrestrial Physics, Garching 85748, Germany}
\address[5]{Higgs Centre for Theoretical Physics, Edinburgh EH9 3FD, UK}
\address[6]{Kavli Institute for Astronomy and Astrophysics, Peking University, Beijing 100871, China}
\address[7]{National Astronomical Observatories, Chinese Academy of Sciences, Beijing 100012, China}

\begin{abstract}
We put into test the idea of replacing dark energy by a vector field against the cosmic microwave background (CMB) observation using the simplest vector-tensor theory, where a massive vector field couples to the Ricci scalar and the Ricci tensor quadratically. First, a remarkable Friedmann-Lema\^{i}tre-Robertson-Walker (FLRW) metric solution that is completely independent of the matter-energy compositions of the universe is found. Second, based on the FLRW solution as well as the perturbation equations, a numerical code calculating the CMB temperature power spectrum is built. We find that though the FLRW solution can mimic the evolution of the universe in the standard $\Lambda$CDM model, the calculated CMB temperature power spectrum shows unavoidable discrepancies from the CMB power spectrum measurements.
\end{abstract}


\end{frontmatter}



\section{Introduction}
\label{sec:intro}

The observation of the cosmic microwave background (CMB) anisotropy has promoted cosmology to a high-precision scientific field of research \cite{Bennett:1996ce,Fixsen:1996nj}. Based on the standard $\Lambda$CDM model in general relativity (GR), CMB data from the Planck mission updated our understanding of the geometry and the compositions of the universe \cite{Planck:2018vyg}. Moreover, the precisely measured CMB anisotropy provides stringent tests for alternative theories of gravity, helping us to judge between dark energy and its substitutions (e.g. see Refs.~\cite{Clifton:2011jh, Bamba:2012cp, Koyama:2015vza, Planck:2015bue, CANTATA:2021ktz}). Motivated by examing the idea of replacing dark energy by an auxiliary vector field rigorously, we build a code to calculate the power spectrum of the CMB temperature anisotropy in an appealing vector-tensor theory, called the bumblebee gravity \cite{Kostelecky:2003fs}, to compare with the standard $\Lambda$CDM result.

The bumblebee gravity theory has the action          
\begin{align}
 S =& \frac{1}{2\kappa} \int d^4x \sqrt{-g} \left( R + \xi_1 B^\mu B^\nu
R_{\mu\nu} + \xi_2 B^\mu B_\mu R \right) 
\nonumber \\
& - \int d^4x \sqrt{-g} \, \left( \frac{1}{4} B^{\mu\nu} B_{\mu\nu} + V \right),
\label{actionB}
\end{align} 
where $B^\mu$ is the auxiliary vector field called the bumblebee field, and $B_{\mu\nu} := D_\mu B_\nu - D_\nu B_\mu$ with $D_\mu$ being the covariant derivative. The theory possesses the two simplest nonminimal coupling terms between the vector field and the curvature quantities, namely the Ricci tensor $R_{\mu\nu}$ and the Ricci scalar $R$, with $\xi_1$ and $\xi_2$ being the coupling constants. The potential $V$ is generally a function of $B^\mu$. In this work, it takes the simplest form
\begin{align}
V = V_1 B^\mu B_\mu ,
\label{massivepot}
\end{align} 
where $V_1$ is a positive constant and represents the square of the effective mass of the bumblebee field. We will mostly use the geometrized units where the gravitational constant $G$ and the speed of light $c$ are set to unity ($\kappa = 8\pi$), though physical quantities are sometimes given in conventional units for a better perception of their sizes. 

A theory similar to the one in Eq.~\rf{actionB} but without the potential $V$ was originally proposed and studied by Hellings and Nordtvedt \cite{Hellings:1973zz}. They calculated coefficients for the parametrized post-Newtonian (PPN) formula and the Friedmann-Lema\^{i}tre-Robertson-Walker (FLRW) metric solution in the theory. Later, Kosteleck\'{y} started to use this theory to introduce spontaneous breaking of local Lorentz symmetry in gravity by adding in the potential $V$ and requiring its minimum to be realized with a nonzero configuration of the bumblebee field \cite{Kostelecky:2003fs}. The theory then received intensive studies in seeking for features and effects of Lorentz-symmetry violation in gravity (e.g. see Refs.~\cite{Bluhm:2004ep,Bertolami:2005bh,Bailey:2006fd,Bluhm:2007bd,Bluhm:2008yt,Kostelecky:2010ze,Ovgun:2018xys}). Recently, interests in this theory as the simplest vector-tensor theory have returned since two branches of solutions of spherical black holes (BHs) were found for the special case of $V=0$ and $\xi_2=0$ \cite{Casana:2017jkc,Xu:2022frb}. One branch of the solutions generalizes the Reissner-Nordstr\"{o}m (RN) BH while the other is a deformation of the Schwarzschild BH due to the accompanied nontrival bumblebee field. Based on the generalized RN BH solutions, the first law of BH thermodynamics has been extended and numerically checked \cite{Mai:2023ggs}, and interesting theoretic bounds on the charge of the BHs have been discovered by investigating the dynamic instabilities of the solutions \cite{Mai:2024lgk}. We point out that the potential which we use in Eq.~\rf{massivepot} simply has its minimum when the bumblebee field vanishes. So it generates no vacuum expectation value of the bumblebee field and cannot cause spontaneous spacetime symmetry breaking. This is usually different in the literature where the potential takes other forms and has its minimum when the bumblebee field is shifted from zero (e.g. see Refs.~\cite{Kostelecky:2003fs}); the nonzero vacuum expectation value of the bumblebee field causes spontaneous spacetime symmetry breaking. 

Appealed by the intriguing BH solutions in the bumblebee theory, we continue the work of Hellings and Nordtvedt by extending their FLRW solution to the case of a massive vector field and carrying out the calculation of the power spectrum of the CMB temperature anisotropy. We find that although the FLRW solution can mimic the expansion of the universe in the standard $\Lambda$CDM scenario, the CMB power spectrum deviates radically from the standard $\Lambda$CDM result at very large scales, disfavoring substituting dark energy with the bumblebee field. It demonstrates how effective the modern CMB observation can be in ruling out the alternative gravity theory that has escaped tests from weak-field observations \cite{Hellings:1973zz,Casana:2017jkc,Xu:2022frb}, BH images \cite{Xu:2022frb,Xu:2023xqh}, and redshift-distance measurements in cosmology \cite{tempxu}. In building our CMB code, we use CAMB \cite{Lewis:1999bs} and CLASS \cite{Blas:2011rf} as important references. Compared to modifications based on them, such as MGCAMB \cite{Zhao:2009fn, Hojjati:2011ix}, we have to derive the approximate solutions at the early universe according to the field equations in the bumblebee theory to set up proper initial conditions for the numerical integration. It is the first public CMB code working for an action-based modified gravity to our knowledge. The code is available at {\url{https://github.com/ryxxastroat/bumblebeecmb}}.

\section{FLRW solution in the bumblebee theory}
\label{sec:II}
We start with the FLRW metric ansatz
\begin{equation}
ds^2 = a^2 \left( -d\eta^2 + \frac{dr^2}{1 - {\cal K}_0 r^2} + r^2 d\Omega^2 \right),
\end{equation} 
where $a$ is the time-depending scale factor and the constant ${\cal K}_0$ represents the current spatial curvature of the universe. The homogeneous and isotropic FLRW metric requires the bumblebee field to take the form of  
\begin{align}
B_\mu = \left( b_\eta, 0, 0, 0 \right),
\end{align}
where $b_\eta$ depends only on the time coordinate $\eta$. 
The energy-momentum tensors for matter and radiation take the usual perfect-fluid form,
\begin{align}
 \left(T_m\right)_{\mu\nu} &= \left( \epsilon_m + p_m \right) \left( u_m \right)_\mu \left( u_m \right)_\nu + p_m g_{\mu\nu}, 
\nonumber \\
 \left(T_r\right)_{\mu\nu} &= \left( \epsilon_r + p_r \right) \left( u_r \right)_\mu \left( u_r \right)_\nu + p_r g_{\mu\nu}, 
\end{align} 
where $\epsilon_A,\, p_A, \left(u_A\right)_\mu$, $A=m,\, r,$ are energy densities, pressures, and four-velocities for matter and radiation. The equations of state for matter and radiation are 
\begin{align}
&& p_m = 0, 
\nonumber \\
&& p_r = \frac{1}{3} \epsilon_r .
\end{align}
The four-velocities are 
\begin{align}
\left( u_m \right)_\mu = \left( u_r \right)_\mu = \left( -a , 0,0,0\right),
\end{align}
to be consistent with the FLRW metric.

With the above setup and using the field equations presented in \ref{app1}, we find two ordinary differential equations (ODEs) for $a$ and $b_\eta$,  
\begin{align}
0 =& -\frac{3 (\xi_1+2 \xi_2) a' b_\eta'b_\eta}{a} + \left( \frac{3 \xi_2 a^{\prime\,2} }{a^2} + \kappa V_1a^2-3 {\cal K}_0 \xi_2\right) b_\eta^2  
\nonumber \\
& + 3 a^{\prime\,2} - \kappa \left(\epsilon_m+\epsilon_r\right) a^4 + 3 {\cal K}_0 a^2,
\nonumber \\
0 =& b_\eta \left[ 3 (\xi_1+2 \xi_2) a'' - 3 \xi_1 \frac{a^{\prime\,2}}{a} + 6 \xi_2 {\cal K}_0 a - 2 \kappa V_1 a^3 \right],
\label{flrweq}
\end{align}
where the primes denote derivatives with respect to the conformal time $\eta$. The first equation is derived from the Einstein field equations in Eq.~\rf{fieldeqs}, and the second equation is from the temporal component of the vector field equation in Eq.~\rf{fieldeqs}. It is also possible to derive the second equation in Eq.~\rf{flrweq} from the Einstein field equations, but it is less straightforward than using the temporal component of the bumblebee field equation. 

Equation~(\ref{flrweq}) interestingly admits two solutions. The first one, with $b_\eta=0$, simply reduces to the GR solution without dark energy. The second one, with $b_\eta \ne 0$, is a nontrivial new FLRW solution and remarkably has an elegant expression for the expansion rate,
\begin{align}
{\cal H} := \frac{a'}{a} = H_0 \sqrt{ \Omega_{V_1} a^2 + \left( 1-\Omega_{V_1}-\Omega_{{\cal K}0} \right) a^\alpha + \Omega_{{\cal K}0}},
\label{flrwsol1}
\end{align} 
where $H_0$ is the Hubble constant, and 
\begin{align}
\Omega_{V_1} =& \frac{2 \tilde V_1}{\xi_1+4\xi_2},
\nonumber \\
\Omega_{{\cal K}0} =& - \frac{{\cal K}_0}{H_0^2} ,
\nonumber \\
\alpha =& - \frac{4\xi_2}{\xi_1+2\xi_2} ,
\end{align}
with $\tilde V_1 = V_1/\epsilon_{\rm cri0}$ and $\epsilon_{\rm cri0} = 3H_0^2/\kappa$ being the current critical energy density of the universe. At the background level, the energy-momentum tensors for matter and radiation are conserved separately so that
\begin{align}
\epsilon_m =& \frac{\epsilon_{m0}}{a^3}, 
\nonumber \\
\epsilon_r =& \frac{\epsilon_{r0}}{a^4}, 
\end{align}
where $\epsilon_{m0}$ and $\epsilon_{r0}$ are the current energy densities of matter and radiation.  
The evolution of $b_\eta$ can then be solved numerically using the first equation in Eq.~(\ref{flrweq}).

The new bumblebee FLRW solution in Eq.~(\ref{flrwsol1}) resembles the standard $\Lambda$CDM solution
\begin{equation}
{\cal H}_{\rm GR} = H_0 \sqrt{ \Omega_{\Lambda 0} a^2 + \Omega_{r0} a^{-2} + \Omega_{m0} a^{-1} + \Omega_{{\cal K}0} },
\label{flrwlcdm}
\end{equation} 
where $\Omega_{\Lambda 0}, \, \Omega_{m0}$ and $\Omega_{r0}$ are the current energy fractions of dark energy, matter and radiation.
But there are two important aspects where they differ. 
\begin{enumerate}
\item There is no cosmological constant in setting up the bumblebee theory; the effect of $V_1$ turns out to mimic the cosmological constant.  

\item Matter and radiation play no role in Eq.~(\ref{flrwsol1}); they are replaced by a new term proportional to $a^\alpha$, where $\alpha$ depends on the ratio between the two coupling constants. 
\end{enumerate}  
The first point is appealing because it provides a possible origin for the cosmological constant or dark energy. The second point, though seems to contradict the doctrine that matter must influence spacetime, one finds that the nonminimal couplings between the bumblebee field and the curvature quantities contribute a negative effective energy density that cancel out the energy densities of matter and radiation. So the expasion rate of the universe depends on the coupling constants $\xi_1, \, \xi_2$, and the effective mass parameter $V_1$ of the bumblebee field, rather than the energy fractions of matter and radiation.

To make the point clear, the bumblebee cosmological model in Eq.~(\ref{flrwsol1}) therefore provides an example where the expansion of the universe is a consequence of spacetime itself interacting with an auxiliary vector field, which possibly emerges from an underlying theory. Matter and radiation, presumably the source of spacetime curvature, become guests visiting the prefixed background universe and cannot influence it at the homogeneous and isotropic background level. The idea is so radically different from the conventional $\Lambda$CDM scenario, yet there are no a priori reasons to exclude it. Predictions from this new bumblebee cosmological model need to be tested against observations to learn more about its validity.

\begin{figure}[h]
 \includegraphics[width=0.98\linewidth]{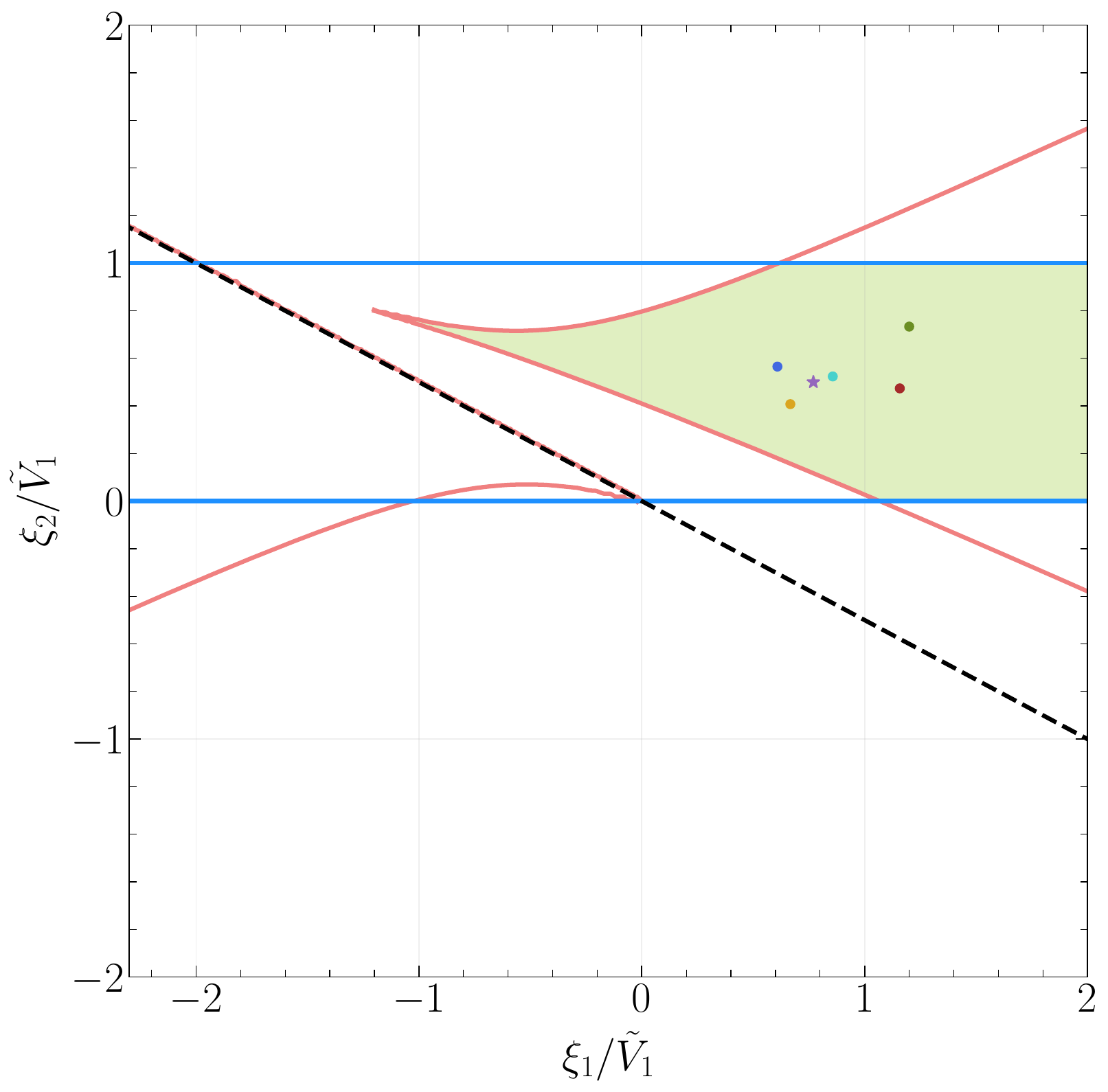}
 \caption{Prior parameter space (the shaded region). The blue lines $\xi_2/\tilde V_1 = 1$ and $\xi_2/\tilde V_1=0$ correspond to $q_0=0$ and $\alpha=0$, while the dashed black line $\xi_1+2\xi_2=0$ corresponds to $q_0 \rightarrow \infty$ and $\alpha \rightarrow \infty$. The red line is the contour for the age of the universe to be $0.68\, H_0^{-1}$, which is about $9.5 \times 10^9$ years with $H_0 \approx 70\, {\rm km/s/Mpc}$. The dots are 5 representative sets of parameters to use for demonstrating the CMB results later (colors consistent with those in Figs.~\ref{fig2} and~\ref{fig4}). The star is the best-fit result for the spatially flat bumblebee model in Ref.~\cite{tempxu}.}
\label{fig1}
\end{figure}

\begin{figure}
 \includegraphics[width=0.98\linewidth]{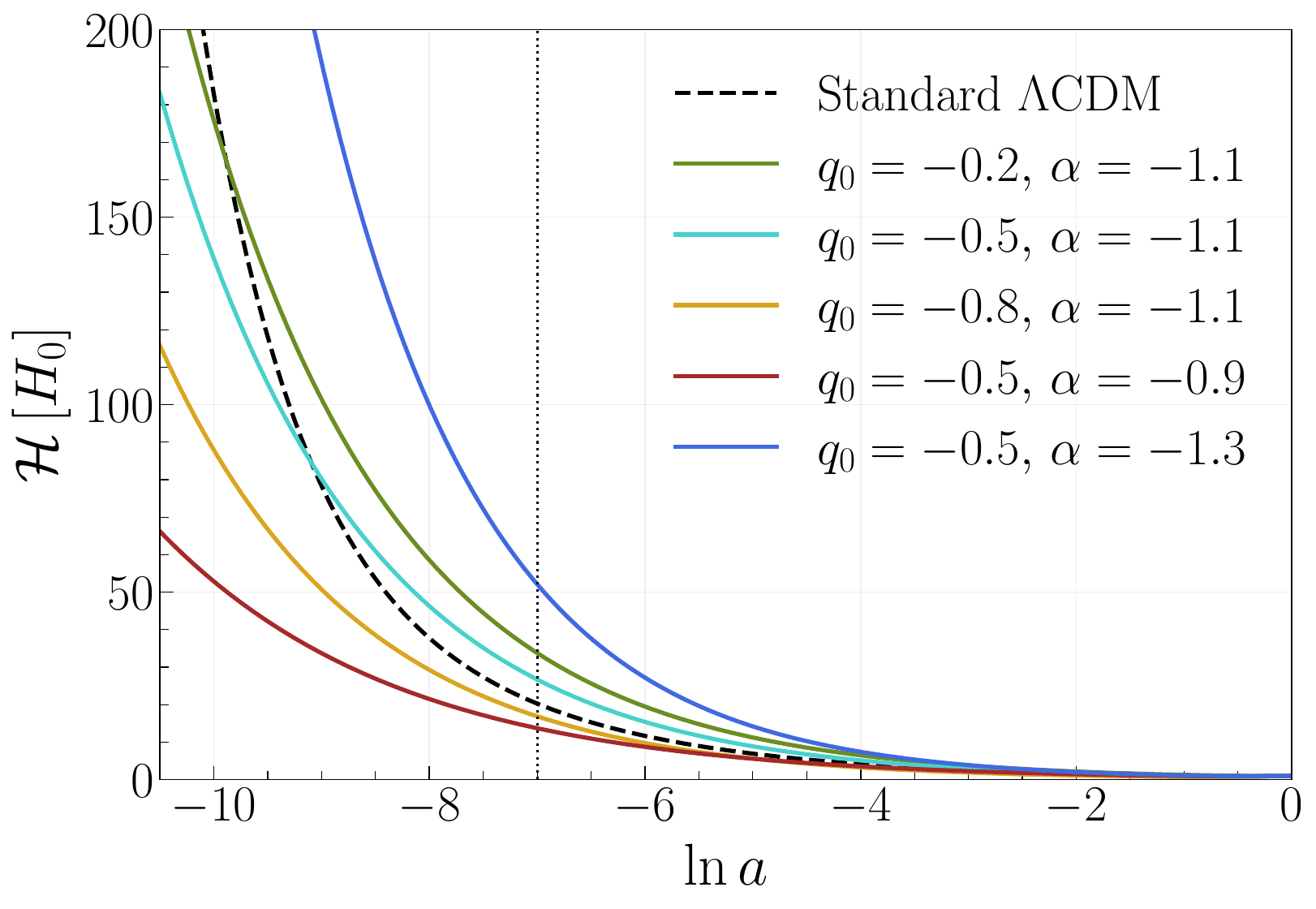}
 \caption{${\cal H}$ vs. $\ln{a}$. The vertical line at $\ln{a}=-7$ marks the epoch of recombination. The parameters of the bumblebee solutions correspond to the dots in Fig.~\ref{fig1}. The standard $\Lambda$CDM solution in GR has parameters $H_0=67.4\, {\rm km/s/Mpc}$, $\Omega_{m0}=0.315$, and $T_{\rm CMB}=2.726\, K$ \cite{Planck:2018vyg}.}
\label{fig2}
\end{figure}

Tests of the FLRW solution in Eq.~(\ref{flrwsol1}) using distance-redshift data from selected standard type Ia supernovae and measurements of baryon acoustic oscillations have been done carefully in a separate work \cite{tempxu}. Sensible best-fit values for the 4 parameters, $\xi_1/\tilde V_1, \, \xi_2/\tilde V_1, \, \Omega_{{\cal K}0},$ and $H_0$, have been obtained. In this work, we concentrate on testing the solution independently using the observation of CMB temperature power spectrum. We will take ${\cal K}_0 =0$, corresponding to a spatially flat cosmological model, to simplify the calculation. For ${\cal K}_0 \ne 0$, we expect the results to be qualitatively the same. This is because the most contribution to the CMB power spectrum happens at recombination when $a \sim 10^{-3}$, so that the behavior of the expansion rate ${\cal H}$ at $a\rightarrow 0$ is crucial. If $\alpha \ge 0$, ${\cal H}$ is finite at $a\rightarrow 0$, which leads to pathological behaviors of the CMB perturbation variables (see \ref{app3}). Therefore, we restrict to $\alpha<0$, in which case a nonzero $\Omega_{{\cal K}0}$ does not affect the $a^{\alpha/2}$ behavior of ${\cal H}$ at $a\rightarrow 0$ and thus is not expected to qualitatively change the CMB results that we are going to show.
To further reduce the parameter space, we require the solution to satisfy two brief observational conditions: (i) The deceleration parameter $q_0=-d\ln{\cal H}/d\ln{a}|_{a=1}$ is negative \cite{SupernovaSearchTeam:1998fmf}. (ii) The age of the universe is greater than $9.5 \times 10^9$ years according to radioactive dating \cite{Cayrel:2001qi}. The resultant parameter space is shown in Fig.~\ref{fig1}. In Fig.~\ref{fig2}, a few examples of the bumblebee FLRW solution with $\Omega_{{\cal K}0}=0$ are plotted. Note that the parameters $q_0$ and $\alpha$ are used in place of the parameters $\xi_1/\tilde V_1$ and $\xi_2/\tilde V_1$, with the relation being
\begin{align}
\frac{\xi_1}{\tilde V_1} = -\frac{2\alpha +4}{\alpha+2q_0}, \quad \frac{\xi_2}{\tilde V_1} = \frac{\alpha}{\alpha+2q_0} .
\label{partrans}
\end{align}

\section{CMB temperature anisotropy in the bumblebee theory}
\label{sec:constraints}
The CMB temperature anisotropy is represented by the correlation function $C\left( \hat {\boldsymbol n} \cdot \hat {\boldsymbol n}' \right) := \langle \Theta(\hat {\boldsymbol n}) \Theta(\hat {\boldsymbol n}') \rangle$, with $\Theta:=\Delta T_{\rm CMB}/T_{\rm CMB}$ being the fractional fluctuation of the CMB temperature and $-\hat {\boldsymbol n}$ representing the unit vector along the received CMB photons' path. On the observational side, the multipoles of $C\left( \hat {\boldsymbol n} \cdot \hat {\boldsymbol n}' \right)$ at the current time $\eta=\eta_0$ and the location of the Earth ${\boldsymbol x} =0$ are measured in Planck's mission \cite{Planck:2018nkj}. On the theoretical side, as presented here for the bumblebee theory and in the literature and textbooks of cosmology for GR (e.g. see Refs.~\cite{Ma:1995ey,Callin:2006qx,Durrer:2008eom,Blas:2011rf,Baumann:2022mni}), the multipoles of $C\left( \hat {\boldsymbol n} \cdot \hat {\boldsymbol n}' \right)$ at $\eta=\eta_0, \, {\boldsymbol x} =0$ are calculated in the following three steps.

First, a truncated system of differential equations governing the evolution of the multipoles of $\Theta$ in the Fourier space needs to be solved. These equations involve perturbations in the spacetime metric and the velocity of baryon matter, on top of the background FLRW description of the universe, so they are solved together with the linear perturbation of the Einstein field equations. The complete set of the equations in the bumblebee theory is presented in \ref{app2}. The appropriate initial condition to solve the set of the equations is derived in \ref{app3}. For explaining our results, we list here the perturbation variables used. 
\begin{align*}
& \Psi: {\rm\ Perturbation\ variable\ in\ } g_{\eta\eta}.
 \\
& \Phi:{\rm\ Scalar\ perturbation\ variable\ in\ } g_{ij}.
 \\
& \delta b_\eta:{\rm\ Perturbation\ of\ } B_{\eta}.
 \\
& \delta b_S:{\rm\ Scalar\ perturbation\ variable\ in\ } B_i.
 \\
& \delta_b:{\rm\ Fractional\ perturbation\ of\ the\ baryon\ density.} 
 \\
& \delta_c:{\rm\ Fractional\ perturbation\ of\ the\ cold\ dark\ matter\ density.}
 \\
& v_b:{\rm\ Scalar\ perturbation\ of\ the\ baryon\ velocity.}
 \\
& v_c:{\rm\ Scalar\ perturbation\ of\ the\ cold\ dark\ matter\ velocity.}
 \\
& \Theta_l:{\rm\ Multipoles\ of\ the\ CMB\ temperature\ fluctuation.}
\end{align*}

Second, the line-of-sight formula \cite{Baumann:2022mni},
\begin{align}
\Theta_l(k) =& \int_0^{\eta_0} d\eta \, \Big[ \left( g \left( \Theta_0 + \Psi \right) + e^{-\tau} \left( \Phi' + \Psi' \right) \right) j_l(k\chi) 
\nonumber \\
& - g v_b \frac{d}{d(k\chi)}j_l(k\chi) \Big] ,
\label{losint}
\end{align}
is used to generate $\Theta_l$ in the Fourier space at $\eta=\eta_0$ for any $l$. In Eq.~(\ref{losint}), $j_l$ are the spherical Bessel functions, $k$ is the magnitude of the wave vector for each Fourier mode, and $\chi = \eta_0-\eta$. The optical depth $\tau$ and the visibility function $g$ depend on the background number density of electrons, which can be calculated using the Peebles equation given the expansion history of the universe \cite{Peebles:1968ja}.

Third, the multipoles of $C\left( \hat {\boldsymbol n} \cdot \hat {\boldsymbol n}' \right)$ at $\eta=\eta_0, \, {\boldsymbol x} =0$ are calculated via
\begin{align}
C_l = 4\pi \int \frac{dk}{k} \Theta_l^2(k) \, \Delta_{\cal R}^2(k) ,
\label{clint}
\end{align}
where $\Delta_{\cal R}^2(k)$ is the initial power spectrum describing the size of the initial perturbations for each Fourier mode. Motivated by the inflation theory \cite{Guth:1980zm,Linde:1981mu}, $\Delta_{\cal R}^2(k)$ takes the form
\begin{align}
\Delta_{\cal R}^2(k) = A_S k^{n_S-1} ,
\end{align}
where the constants $A_S$ and $n_S$ are determined by fitting the theoretically calculated $C_l$ to the observational data.

Following the three steps, we have developed a code using Wolfram Mathematica (solving the equations in step one) and Python (calculating numerical integrals in steps two and three) to compute the multipoles $C_l$ in the bumblebee theory. The code is available at {\url{https://github.com/ryxxastroat/bumblebeecmb}}. 
Here we present our numerical results at three levels accordingly: (i) evolution of the perturbation variables, (ii) $\Theta_l$ at $\eta=\eta_0$ as functions of $k$, and (iii) $C_l$ changing with $l$.

\begin{figure}
 \includegraphics[width=0.98\linewidth]{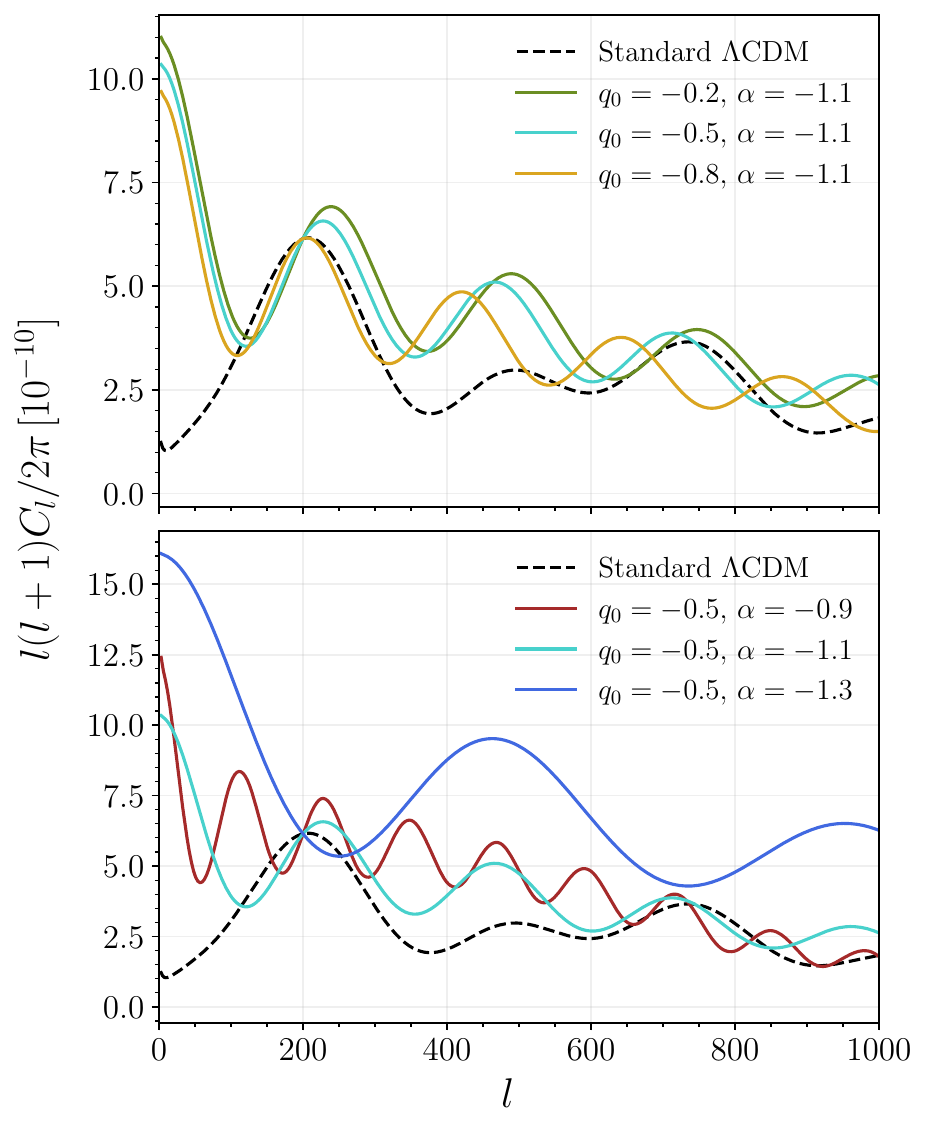}
 \caption{Representative results of the CMB temperature power spectrum calculated in the bumblebee theory. Upper panel: Comparing the results when changing the parameter $q_0$. Lower panel: Comparing the results when changing the parameter $\alpha$. The bumblebee cosmological model has the conventional parameters $H_0=70\, {\rm km/s/Mpc}$, $\Omega_{b0}=0.05$, $\Omega_{c0}=0$, $T_{\rm CMB}=2.7\, K$ and $n_s=1$. The standard $\Lambda$CDM result, which is calculated using the parameters $H_0=67.4\, {\rm km/s/Mpc}$, $\Omega_{b0}=0.0493$, $\Omega_{c0}=0.2657$, $T_{\rm CMB}=2.726\, K$, and $n_s=0.965$ \cite{Planck:2018vyg}, is plotted in both panels for comparison. We have normalized the bumblebee results to match the standard $\Lambda$CDM result at $l=200$. 
}
\label{fig4}
\end{figure}

We start with the final results of $C_l$ in Fig.~\ref{fig4}. The plot shows the CMB power spectrum in the bumblebee gravity calculated for representative values of the parameters $q_0$ and $\alpha$. An impressive observation is that the parameters $q_0$ and $\alpha$ each determines one important feature in the power spectrum plot: the peaks of the power spectrum shift to right as $q_0$ increases while the distances between two consecutive peaks shrink as $\alpha$ increases. This observation makes it possible to adjust $q_0$ and $\alpha$ to more or less fit the standard $\Lambda$CDM result at large $l$. But there is then the fatal discrepency at small $l$ between the bumblebee results and the standard $C_l$ curve: The unexpected raise of the power spectrum as $l$ goes from about $100$ to $1$ in the bumblebee results rules out any possibility of a sensible match to the standard CMB power spectrum.

Before shedding light on the cause of the discrepency, let us point out that the results are found to be independent of the parameter $V_1$ as long as $\xi_1/\tilde V_1$ and $\xi_2/\tilde V_1$ are fixed. 
So $q_0$ and $\alpha$, equivalently $\xi_1/\tilde V_1$ and $\xi_2/\tilde V_1$, are the only two parameters to adjust in our model. There are the conventional parameters $H_0, \, T_{\rm CMB}$, and the fraction of baryonic matter $\Omega_{b0}$, but they are more or less fixed by observations \cite{HST:2000azd,Fixsen:2009ug,2dFGRS:2005yhx}. Small changes of them alter the results insignificantly. 
Then there is the fraction of dark matter $\Omega_{c0}$, but it turns out to have no significant effect at all. In fact, we have set it to zero. The initial power spectrum index $n_S$ changes the overall tilt of the calculated power spectrum in a predictable way, so the approximation $n_S=1$ has been used.

\begin{figure*}
 \includegraphics[width=0.98\linewidth]{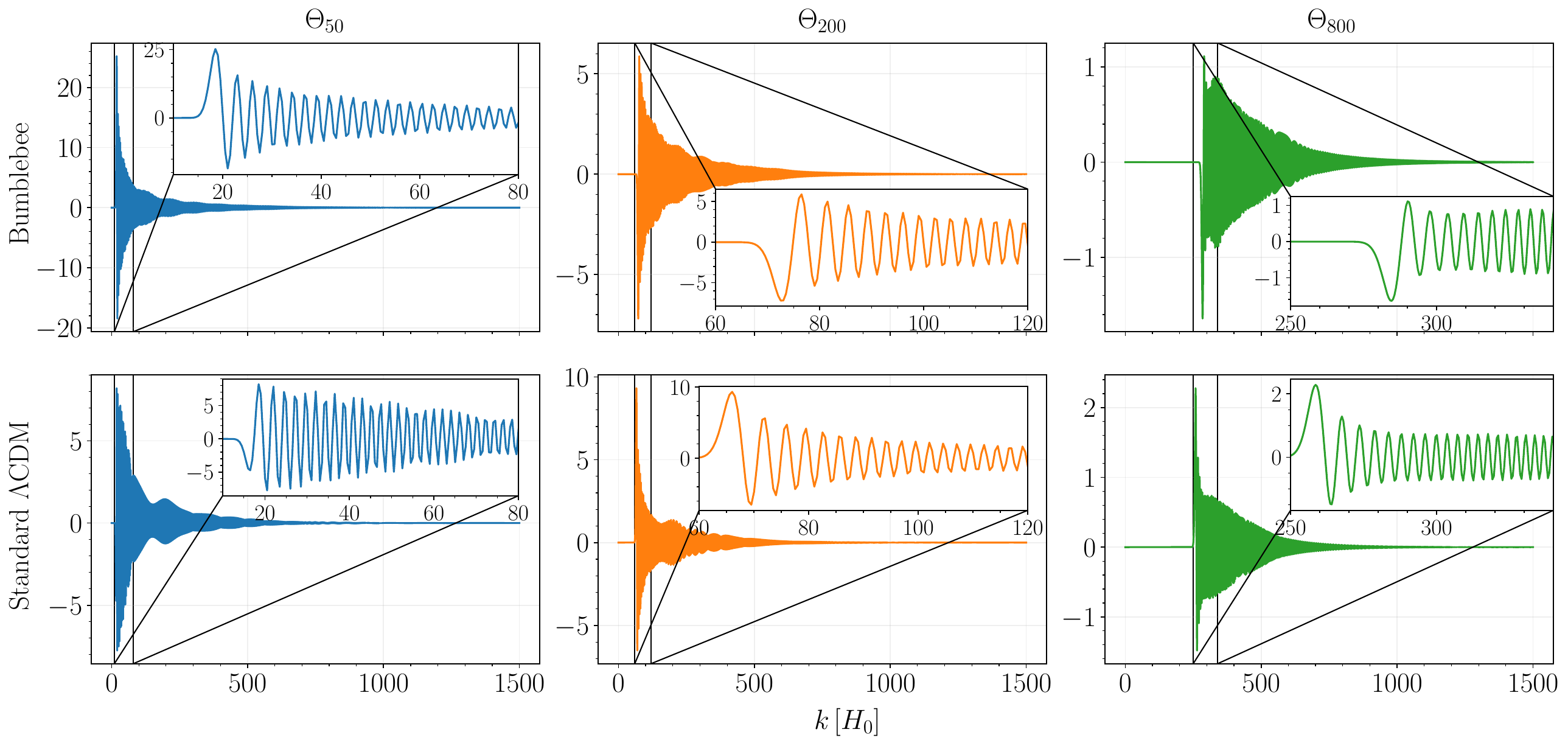}
 \caption{Examples of $\Theta_l$ calculated using Eq.~(\ref{losint}) in the bumblebee cosmological model compared with the standard $\Lambda$CDM results. The parameters for the bumblebee cosmological model are set to be $H_0=70\, {\rm km/s/Mpc}$, $\Omega_{b0}=0.05$, $\Omega_{c0}=0$, $T_{\rm CMB}=2.7\, K$, $n_s=1$, $q_0=-0.5$, and $\alpha=-1.1$. The parameters of the standard $\Lambda$CDM model are $H_0=67.4\, {\rm km/s/Mpc}$, $\Omega_{b0}=0.0493$, $\Omega_{c0}=0.2657$, $T_{\rm CMB}=2.726\, K$, and $n_s=0.965$ \cite{Planck:2018vyg}.     }
\label{fig31}
\end{figure*}

\begin{figure*}
 \includegraphics[width=0.98\linewidth]{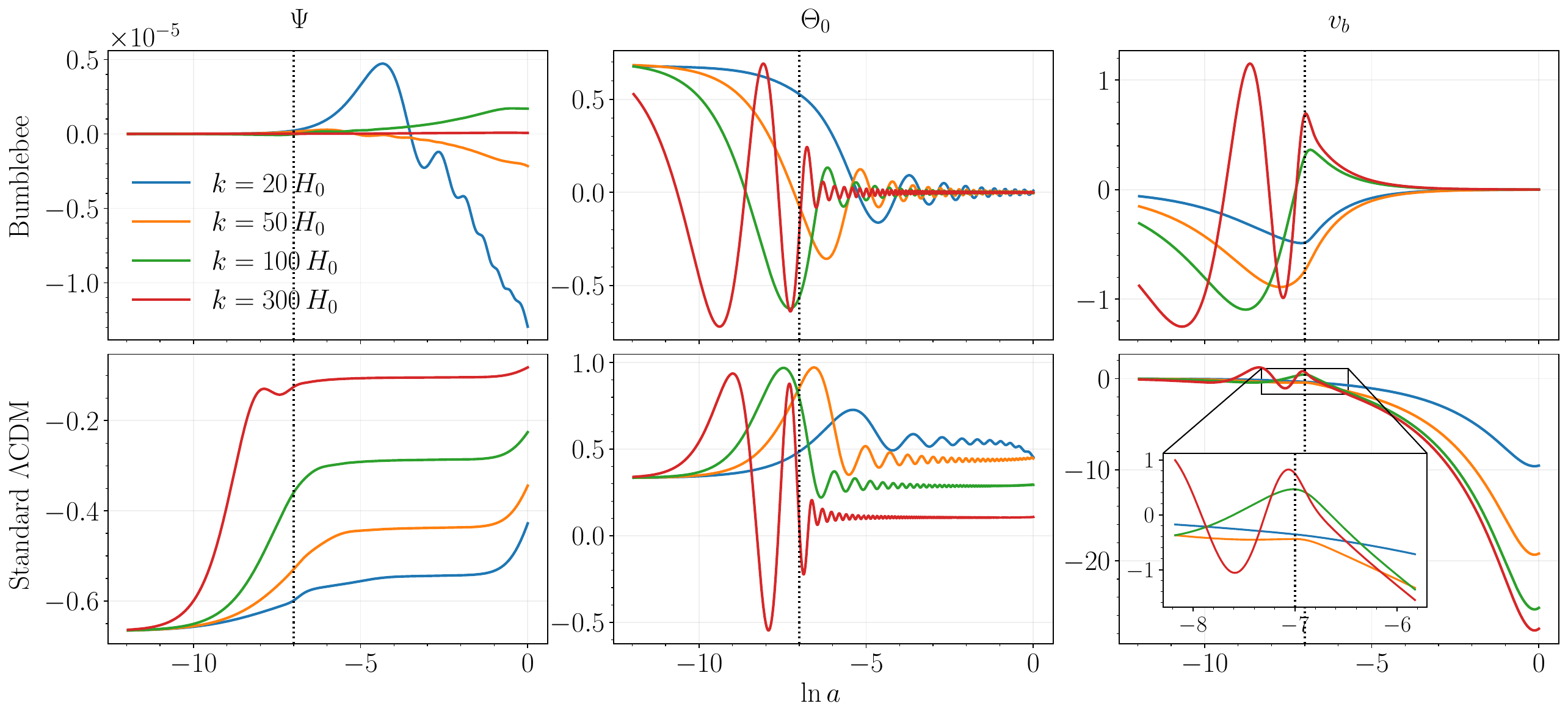}
 \caption{Typical numerical solutions for the representative perturbation variables $\Psi, \ \Theta_0$, and $v_b$. Four Fourier modes with different values of $k$ are shown. The vertical line at $\ln{a}=-7$ marks the epoch of recombination. The parameters for the bumblebee cosmological model are set to be $H_0=70\, {\rm km/s/Mpc}$, $\Omega_{b0}=0.05$, $\Omega_{c0}=0$, $T_{\rm CMB}=2.7\, K$, $n_s=1$, $q_0=-0.5$, and $\alpha=-1.1$. The parameters of the standard $\Lambda$CDM model are $H_0=67.4\, {\rm km/s/Mpc}$, $\Omega_{b0}=0.0493$, $\Omega_{c0}=0.2657$, $T_{\rm CMB}=2.726\, K$, and $n_s=0.965$ \cite{Planck:2018vyg}. Note that the $10^{-5}$ factor is only for $\Psi$ in the bumblebee model.  }
\label{fig3}
\end{figure*}

Now let us look at the middle-level results, $\Theta_l$ at $\eta=\eta_0$ as functions of $k$. Typical examples of $\Theta_l$ are presented in Fig.~\ref{fig31}. They demonstrate that for large $l$, $\Theta_l$ in the bumblebee model and in the standard $\Lambda$CDM model have comparable sizes, while for small $l$, $\Theta_l$ in the bumblebee model are visibly larger. Let us point out that the global maxima of $\Theta_l$ in Fig.~\ref{fig31} appear at $k \sim l/\eta_0$ with $\eta_0 \sim 3 H_0^{-1}$, because that is where the integrand in Eq.~(\ref{losint}) contributes most. To be specific, the spherical Bessel functions $j_l(x)$ and their derivatives have global maxima at $x \sim l$, and the visibility function $g$ has a sharp peak at the epoch of recombination when the conformal time is $\eta=\eta_*\approx 0$. So they together pick up the integrand at $k\sim l/(\eta_0-\eta_*) \sim l/\eta_0$. Note that the term $e^{-\tau}\left( \Phi'+\Psi'\right)$ is much smaller than the other terms, so it can be neglected when undertanding the qualitative features in the results of $\Theta_l$.  

Lastly, to explain why $\Theta_l$ at small $l$ in the bumblebee model are larger than those in the standard $\Lambda$CDM model, we come to the evolution of the perturbation variables. In Fig.~\ref{fig3}, typical solutions of the perturbation variables used in Eq.~(\ref{losint}) to calculate $\Theta_l$ are shown. The biggest difference between the bumblebee results and the standard $\Lambda$CDM results is the evolution of the metric perturbation variable $\Psi$ ($\Phi$ has a similar behavior to $\Psi$). In the bumblebee model, $\Psi$ is incredibly small. This is obtained from numerically solving the complete set of the equations governing the evolution of the perturbation variables. The key of the numerical integration of the system is finding out the initial conditions. The detailed derivation is presented in \ref{app3}. There we find that the modified Einstein field equations accompanied with the vector field equation in the bumblebee theory lead to $\Psi \sim \Phi \sim a^{-\alpha/2}$ at $a\rightarrow 0$. It is vastly different from the GR case where $\Psi$ and $\Phi$ are approximately constant at $a \rightarrow 0$. As $\alpha$ is negative under our consideration, $\Psi$ and $\Phi$ are condemned to be small. 

So we can explain the size discrepancy between $\Theta_l$ in the bumblebee model and in the standard $\Lambda$CDM model at small $l$ now.  
Neglecting the relatively small term $e^{-\tau}\left( \Psi'+\Phi'\right)$, the main contribution of the integrand in Eq.~(\ref{losint}) comes from $\Theta_0+\Psi$ and $v_b$ around the recombination time $\eta=\eta_*$ ($\ln{a} \sim -7$ at recombination) where the visibility function $g$ peaks sharply. 
From Fig.~\ref{fig3}, the magnitudes of $\Theta_0$ and $v_b$ in the bumblebee model and in the standard $\Lambda$CDM model are comparable at recombination. But in the bumblebee model, $\Psi$ is extremely small so $\Theta_0+\Psi \approx \Theta_0$. In the standard $\Lambda$CDM model, the contribution from $\Psi$ at recombination is significant for small-$k$ Fourier modes. It cancels $\Theta_0$ so that the standard $\Lambda$CDM model has smaller $\Theta_0+\Psi$ at recombination compared to the bumblebee model. The smaller $k$ is, the more significant for the cancellation. Multiplied by the spherical Bessel functions and their derivatives, the global maximum of the integrand in Eq.~(\ref{losint}) satisfies $l \sim k\eta_0$. So small-$k$ Fourier modes are correspondingly the dominant contribution for $\Theta_l$ with small $l$. Therefore, $\Theta_l$ in the standard $\Lambda$CDM model have smaller magnitudes than those in the bumblebee model for small $l$.

\section{Conclusions}
\label{sec:disc}
First, an interesting FLRW solution in the bumblebee gravitational theory has been found. The solution depicts the expansion of the universe completely through the coupling constants and the mass of the bumblebee vector field, while the compositions of the universe, namely matter and radiation, play no role in the FLRW solution. This makes the bumblebee cosmology conceptually distinctive from the $\Lambda$CDM cosmology where it is the compositions of the universe that determine the expansion rate. 
Despite the profound difference from the standard cosmological model, a decently large parameter space in the bumblebee theory has been found to survive from the two preliminary requirements from astrophysical observations: (i) currently the expansion of the universe is accelerating \cite{SupernovaSearchTeam:1998fmf}, and (ii) the age of the universe needs to be at least greater than $9.5 \times 10^9$ years \cite{Cayrel:2001qi}. 

Then, to further explore the possibility for the bumblebee cosmology to replace the standard $\Lambda$CDM model, we have calculated the power spectrum of the CMB temperature in the bumblebee cosmological model. We find that when $l$ approaches zero, $C_l$ calculated in the bumblebee cosmological model unavoidably increases as $l$ decreases as shown in Fig.~\ref{fig4}, which forbids any sensible match between the bumblebee $C_l$ curves and the standard $\Lambda$CDM one. Tracing this disastrous discrepancy between the bumblebee cosmological model and the standard $\Lambda$CDM model for $C_l$ at small $l$ in each step of our calculation, we find that it is ultimately because the metric perturbation is extremely small in the bumblebee cosmological model. While the metric perturbation is sourced by the perturbations of the matter density and the CMB photon density in GR, the existence of the perturbation of the bumblebee field in the bumblebee theory seems to cause an effectively negative energy density that balances the perturbations of the matter density and the CMB photon density to an extraordinary accuracy, so that the total source for the metric perturbation almost vanishes in the bumblebee theory.       

By building our own CMB code from scratch and analyzing the calculated results, we have demonstrated how effective the CMB observation can be in testing action-based gravitational theories. Our code contains only the minimal ingredients necessary for calculating the CMB power spectrum by sacrificing the precison (about $10\%$ relative error in $C_l$, estimated by applying our code to the $\Lambda$CDM model; see \ref{app4}) and the speed. So it is much more transparent to be adapted for other modified gravity theories, compared to CAMB \cite{Lewis:1999bs} and CLASS \cite{Blas:2011rf}. In addition, existing CMB code dealing with modified gravity theories, such as MGCAMB \cite{Zhao:2009fn, Hojjati:2011ix}, requires the theories to admit parametrizations to fit in its framework, which is usually not the case for action-based gravity theories. Our code then serves as an adequate starting point to get a grasp on the CMB power spectrum in numerous such theories, providing quick tailored CMB tests against them complementary to other celebrated tests, such as the solar-system weak-field observations, pulsar timing and pulse profiles, gravitational waves from compact binary coalescences, and black hole images.

In conclusion, the bumblebee cosmological model studied in this work is excluded by the CMB observation, though it has an interesting FLRW solution that does not depend on the compositions of the universe and needs neither dark energy nor dark matter. We point out that a generalization of our model by adding in spatial components of the bumblebee field and replacing the FLRW metric with the Bianchi I model might cure the issue for the small multipoles encountered in our work. Maluf and Neves have studied the case for a slightly different bumblebee theory \cite{Maluf:2021lwh}. For our theory, we recently notice that De Felice and Hell have carried out a detailed mode analysis for cosmological perturbations \cite{DeFelice:2025ykh}. A comprehensive study on the behavior of the CMB power spectrum in the Bianchi I spacetime and the development of the CMB code in this generalized spacetime are interesting but challenging directions for future investigation.

\section*{Acknowledgements}
We thank the anonymous referees for constructive comments and suggestions. We are grateful to Prof. Cheng Zhao for insightful discussions. RX is greatful to Prof. Antony Lewis for the valuable help on the use of CAMB. This work was supported by the National Natural Science Foundation of China (Grants No.~12405070), the China Postdoctoral Science Foundation (2023M741999, GZC20240872), and the high-performance computing cluster in the Astronomy Department at Tsinghua University.

\appendix
\section{Field equations in the bumblebee theory}
\label{app1}

The field equations given by the action in Eq.~(\ref{actionB}) are
\begin{align}
& G_{\mu\nu} = \kappa T_{\mu\nu} + \kappa \left(T_{B}\right)_{\mu\nu} + \xi_1 \left(T_{B1}\right)_{\mu\nu} + \xi_2 \left(T_{B2}\right)_{\mu\nu} , 
\nonumber \\
& D^\mu B_{\mu\nu} - \frac{dV}{dB^\nu} + \frac{\xi_1}{\kappa} B^\mu R_{\mu\nu} + \frac{\xi_2}{\kappa} B_\nu R = 0 ,
\label{fieldeqs}
\end{align}
where $T_{\mu\nu}$ consists of usual energy-momentum tensors for matter and radiation, the energy-momentum tensor for the bumblebee vector field is 
\begin{align}
\left(T_{B}\right)_{\mu\nu} =& B_{\mu\lambda}B_\nu^{\phantom\nu\lambda} - g_{\mu\nu} \left( \frac{1}{4} B^{\alpha\beta} B_{\alpha\beta} + V \right) 
\nonumber \\
& + 2 B_\mu B_\nu \frac{dV}{d(B^\lambda B_\lambda)} ,
\end{align}
and the contributions due to the couplings between the bumblebee field and the spacetime curvature are 
\begin{align}
 \left(T_{B1}\right)_{\mu\nu} =& \frac{1}{2} g_{\mu\nu} B^\alpha B^\beta R_{\alpha\beta} - B_\mu B_\lambda R_\nu^{\phantom\nu \lambda} - B_\nu B_\lambda R_\mu^{\phantom\mu \lambda} 
\nonumber \\
& + \frac{1}{2} \Big[ D_\kappa D_\mu \left( B^\kappa B_\nu \right) + D_\kappa D_\nu \left( B_\mu B^\kappa \right) 
\nonumber \\
& - \Box_g \left( B_\mu B_\nu \right) - g_{{\mu\nu}} D_\alpha D_\beta \left( B^\alpha B^\beta \right) \Big], 
\nonumber \\
 \left(T_{B2}\right)_{\mu\nu} =& -B^\lambda B_\lambda G_{\mu\nu} - B_\mu B_\nu R + D_\mu D_\nu \left( B^\lambda B_\lambda \right) 
\nonumber \\
&  - g_{\mu\nu} \Box_g \left( B^\lambda B_\lambda \right) ,
\end{align}
with $\Box_g = D_\mu D^\mu$. 

Besides the field equations, the energy-momentum conservation equation for matter and radiation,
\begin{align}
D^\mu T_{\mu\nu} = 0 ,
\label{coneqtot}
\end{align} 
is useful in calculations. Note that Eq.~(\ref{coneqtot}) can be derived from the field equations. In fact, the covariant divergences of $\left( T_B \right)_{\mu\nu}, \, \left( T_{B1} \right)_{\mu\nu}$ and $\left( T_{B2} \right)_{\mu\nu}$ are found to be
\begin{align}
& D^\mu \left( T_B \right)_{\mu\nu} = \left(B_\nu^{\phantom\nu\lambda} - B_\nu D^\lambda \right) \left( D^\kappa B_{\kappa\lambda} - \frac{dV}{dB^\lambda} \right), 
\nonumber \\
& D^\mu \left( T_{B1} \right)_{\mu\nu} = \left( B_\nu^{\phantom\nu\lambda} - B_\nu D^\lambda \right) \left( B^\kappa R_{\lambda\kappa} \right) ,
\nonumber \\
& D^\mu \left( T_{B2} \right)_{\mu\nu} = \left( B_\nu^{\phantom\nu\lambda} - B_\nu D^\lambda \right) \left( B_\lambda R \right) ,
\end{align}
so that the conservation of the total of them is guaranteed by the vector field equation in Eq.~(\ref{fieldeqs}). Together with the identity $D^\mu G_{\mu\nu} =0$, Eq.~(\ref{coneqtot}) is just a consequence of the Einstein field equations.

\section{Linear perturbation equations in the bumblebee theory}
\label{app2}
The setup and equations of the linear perturbation theory used for the standard $\Lambda$CDM cosmology can be found in the literature (e.g. see Refs.~\cite{Ma:1995ey,Baumann:2022mni}). For using with the bumblebee theory, we need to add in the bumblebee field and derive the perturbation equations from Eq.~(\ref{fieldeqs}). To begin with, the following provides notations of the scalar-vector-tensor (SVT) decompositions of the perturbation variables used in our CMB code. Note that we restrict to the spatially flat FLRW background solution.   

First, the metric in the conformal coordinates $(\eta, x, y, z)$ is
\begin{align}
g_{\mu\nu} = a^2 (\eta_{\mu\nu} + h_{\mu\nu}),
\end{align}
where $h_{\mu\nu}$ is the perturbation on top of a Minkowski metric $\eta_{\mu\nu}$. The SVT decompositions of $h_{\eta i}$ and $h_{ij}$ are    
\begin{align}
 h_{\eta i} =& \partial_i h_S + \left( h_{V} \right)_i ,
\nonumber \\
 h_{ij} =& \left(h_{TT}\right)_{ij} + \partial_i \left( h_{T}\right)_j + \partial_j \left( h_{T}\right)_i  
\nonumber \\
& + \frac{1}{2} \left( \delta_{ij} \nabla^2 - \partial_i \partial_j \right) h_{1} 
\nonumber \\
& + \frac{1}{2} \left( 3\partial_i\partial_j - \delta_{ij} \nabla^2 \right) h_{2} ,
\end{align}
where $\nabla^2=\delta^{kl}\partial_k \partial_l$. The vector parts $\left( h_{V} \right)_i,\, \left( h_{T} \right)_i$ and the tensor part $\left(h_{TT}\right)_{ij}$ satisfy the constraints 
\begin{align}
& \partial_i \left( h_{V} \right)_i = \partial_i \left( h_{T} \right)_i = \partial_j \left(h_{TT}\right)_{ij} = 0, 
\nonumber \\
& \delta^{ij} \left(h_{TT}\right)_{ij}=0 .
\end{align}

Second, the conventional energy-momentum tensor $T_{\mu\nu}$ in our bumblebee cosmology model includes three ingredients,
\begin{align}
\left(T_b\right)_{\mu\nu} =& \left( \epsilon_b + \delta \epsilon_b \right) \left(u_b\right)_\mu \left(u_b\right)_\nu ,
\nonumber \\
\left(T_c\right)_{\mu\nu} =& \left( \epsilon_c + \delta \epsilon_c \right) \left(u_c\right)_\mu \left(u_c\right)_\nu ,
\nonumber \\
\left(T_\gamma\right)_{\mu\nu} =& \frac{4}{3} \left( \epsilon_\gamma + \delta \epsilon_\gamma \right) \left( u_\gamma \right)_\mu \left( u_\gamma \right)_\nu 
\nonumber \\
& + \frac{1}{3} \left( \epsilon_\gamma + \delta \epsilon_\gamma \right) g_{\mu\nu}  + \Pi_{\mu\nu} , 
\end{align}
where the subscripts $b,\, c$ and $\gamma$ represent baryon, cold dark matter and photon. The background energy densities of the ingredients are denoted as $\epsilon_b, \, \epsilon_c$ and $\epsilon_\gamma$ while their perturbations are $\delta \epsilon_b, \, \delta \epsilon_c$ and $\delta \epsilon_\gamma$. We have used the approximations that matter is pressureless and that photons’ pressure is a third of their energy density. The four-velocities take the form
\begin{align}
& \left( u_b \right)_\eta = \left( u_c \right)_\eta = \left( u_\gamma \right)_\eta = -a \left( 1-\frac{1}{2}h_{\eta\eta} \right),
\nonumber\\
& \left( u_b \right)_i = \partial_i \, \delta u_{bS} + \left(\delta u_{bV} \right)_i, 
\nonumber\\
& \left( u_c \right)_i = \partial_i \, \delta u_{cS} + \left(\delta u_{cV} \right)_i,
\nonumber\\
& \left( u_\gamma \right)_i = \partial_i \, \delta u_{\gamma S} + \left(\delta u_{\gamma V} \right)_i,
\end{align}
up to the linear order of the perturbations, with the constraints 
\begin{align}
\partial_i \left(\delta u_{bV} \right)_i = \partial_i \left(\delta u_{cV} \right)_i = \partial_i \left(\delta u_{\gamma V} \right)_i = 0 .
\end{align}
The last term in $(T_\gamma)_{\mu\nu}$ is a higher-order contribution to the energy-momentum tensor of the photons beyond the perfect-fluid description due to the CMB temperature fluctuation. It takes the form
\begin{align}
\Pi_{\eta i} =& 0 ,
\nonumber \\
\Pi_{ij} =& \left(\Pi_{TT}\right)_{ij} + \partial_i \left( \Pi_{T}\right)_j + \partial_j \left( \Pi_{T}\right)_i  
\nonumber \\
& + \frac{1}{2} \left( 3\partial_i\partial_j - \delta_{ij} \nabla^2 \right) \Pi ,
\end{align} 
with the constraints 
\begin{align}
& \partial_i \left(\Pi_{T}\right)_i = \partial_j \left(\Pi_{TT}\right)_{ij} = 0 ,
\nonumber \\
& \delta^{ij} \left(\Pi_{TT}\right)_{ij} = 0.
\end{align}

Third, the bumblebee field consists of the background part $B^{(0)}_\mu = \left( b_\eta, 0,0,0\right)$ and the perturbation part $B^{(1)}_\mu$ that has an SVT decomposition
\begin{align}
B^{(1)}_\eta =& \delta b_\eta, 
\nonumber \\
B^{(1)}_i =& \partial_i \, \delta b_{S} + \left(\delta b_{V} \right)_i .
\end{align}
The background bumblebee field has only the temporal component to be consistent with the homogeneous and isotropic FLRW spacetime. The perturbation of the bumblebee field has two scalars $\delta b_\eta$ and $\delta b_S$, and a vector $\left( \delta b_V\right)_i$ subject to the constraint
\begin{align}
\partial_i \left(\delta b_{V} \right)_i = 0 .
\end{align} 

Now with the above setup, the field equations in Eq.~(\ref{fieldeqs}) simplify to Eq.~(\ref{flrweq}) at the zeroth order of perturbation with $\epsilon_m = \epsilon_b+\epsilon_c$ and $\epsilon_r=\epsilon_\gamma$; at the linear order of perturbation, they produce three sets of equations for the scalar, the vector, and the tensor variables. Each set of equations does not mix with others. For the calculation of the CMB temperature anisotropy, we only need the set of equations for the scalar variables. Fixing the freedom of the coordinates at the perturbation level by choosing the Newtonian gauge, 
\begin{align}
& h_{\eta\eta} = -2 \Psi , 
\nonumber \\
& h_S = 0, 
\nonumber \\
& \frac{1}{3} \nabla^2 h_1 = \nabla^2 h_2 = -2 \Phi , 
\end{align}
so that the metric perturbation takes the simple form 
\begin{align}
ds^2 = a^2 \left[ -(1+2\Psi) d\eta^2 + \left( 1-2\Phi \right) \delta_{ij} dx^i dx^j \right],
\end{align}    
we find 4 equations for the 4 scalar variables $\Psi, \, \Phi, \, \delta b_\eta$ and $\delta b_S$,
\begin{align}
0 =& \left( 1 - \frac{(\xi_1 + \xi_2)b_\eta^2}{a^2} \right) \Phi' - \frac{(\xi_1 + 2\xi_2)b_\eta^2}{2a^2} \Psi' 
\nonumber \\
& + \left( {\cal H} + \frac{3\xi_2 b_\eta^2 {\cal H} }{a^2} - \frac{3(\xi_1 + 2\xi_2)b_\eta b_\eta'}{2a^2} \right) \Psi  
\nonumber \\
& + \frac{(\xi_1 + 2\xi_2)b_\eta}{2a^2} \delta b_\eta' + \frac{-{\cal H} (\xi_1+6\xi_2) b_\eta + (\xi_1+2\xi_2) b_\eta'}{2a^2} \delta b_\eta
\nonumber \\
&  + \frac{\kappa}{2} a \left( \epsilon_b u_{bS} + \epsilon_c u_{cS} + \frac{4}{3}\epsilon_\gamma u_{\gamma S} \right) ,
\nonumber \\
0 =& \Phi-\Psi - \frac{\xi_2 b_\eta^2}{a^2} (\Phi+\Psi) + \frac{2\xi_2 b_\eta}{a^2} \delta b_\eta + \frac{\xi_1 b_\eta'}{a^2} \delta b_S
\nonumber \\
& + \frac{\xi_1 b_\eta}{a^2} \delta b_S' - \frac{3}{2}\kappa \Pi ,
\nonumber \\
0 =& \left( 3 {\cal H} - \frac{3{\cal H} \xi_1 b_\eta^2 }{2a^2} - \frac{3(\xi_1 + 2\xi_2)b_\eta' b_\eta}{2a^2} \right) \Phi' 
\nonumber \\
& - \frac{3{\cal H} (\xi_1+2\xi_2) b_\eta^2}{2a^2} \Psi' - \left( 1 - \frac{\xi_2 b_\eta^2}{a^2} \right) \nabla^2 \Phi 
\nonumber \\
& + \frac{3{\cal H} (\xi_1+2\xi_2) b_\eta}{2a^2} \delta b_\eta' - \frac{\kappa b_\eta}{2a^2} \nabla^2 \delta b_S'
\nonumber \\
&  +  \frac{b_\eta \left(-6 {\cal H} (\xi_1+2 \xi_2) b_\eta'+6 {\cal H}^2 \xi_2 b_\eta + b_\eta (\xi_1+2 \xi_2) \nabla^2 \right)}{2 a^2} \Psi 
\nonumber \\
&  + a^2 \kappa \left(\epsilon_b+\epsilon_c+\epsilon_\gamma\right) \Psi + \frac{\kappa}{2} a^2 \left(\delta \epsilon_b+\delta\epsilon_c+\delta\epsilon_\gamma\right)
\nonumber \\
& + \frac{b_\eta \left(-2 \kappa a^2 V_1 - 6 {\cal H}^2 \xi_2 + \left(\kappa - \xi_1-2\xi_2\right) \nabla^2 \right)}{2 a^2}  \delta b_\eta
\nonumber \\
& + \frac{3 {\cal H} (\xi_1+2 \xi_2) b_\eta'}{2 a^2} \delta b_\eta - \frac{{\cal H} \xi_1 b_\eta \nabla^2}{a^2} \delta b_S  ,
\nonumber \\
0 =& -\frac{2 \xi_1 \left(3 {\cal H}^2 (\xi_1+4 \xi_2)-2 a^2 \kappa V_1 \right) \delta b_S}{3 \kappa (\xi_1+2 \xi_2)}
\nonumber \\
& +\frac{2 \xi_1 b_\eta \left({\cal H} \Psi+\Phi'\right)}{\kappa}-\delta b_\eta'+\delta b_S'' ,
\label{fieldeqs1}
\end{align}
where the primes denote derivatives with respect to $\eta$ and ${\cal H} = a'/a$. The quadratic potential $V = V_1 B^\mu B_\mu$ has been used. 

To complete the set of equations for all the scalar variables, we also need the energy-momentum conservation equations and the Boltzmann equation to describe the evolution of the scalar variables in the matter and photon sector. Because the bumblebee field has no direct interaction with matter and photons, these equations for the perturbation variables of matter and photon are the same as those in GR, which are \cite{Baumann:2022mni}
\begin{align}
\delta_b' =& \, 3  \Phi' + k  v_b ,
\nonumber \\
v_b' =& - k \Psi - {\cal H} v_b - \frac{4 \epsilon_{\gamma}}{3\epsilon_b} \Gamma \left( 3\Theta_1 + v_b \right)  ,
\nonumber \\
\delta_c' =& \, 3  \Phi' + k  v_c ,
\nonumber \\
v_c' =& - k \Psi - {\cal H} v_c ,
\nonumber \\
\Theta_0' =& \, \Phi'  - k \Theta_1 ,
\nonumber \\
\Theta_1' =& \, \frac{k}{3} \Theta_0 + \frac{k}{3} \Psi - \frac{2k}{3} \Theta_2 -\Gamma\left( \Theta_1 + \frac{1}{3} v_b \right) ,
\nonumber \\
\Theta_l' =& \, - \frac{k}{2l+1} \left[ (l+1)\Theta_{l+1} - l \Theta_{l-1} \right] 
\nonumber \\
&  - \Gamma \left( 1-\frac{1}{10}\delta_{l2} \right) \Theta_l  , \quad l \ge 2 ,
\label{boltzeq}
\end{align} 
where $\Theta_l$ are the multipoles of the CMB temperature fluctuation $\Theta:= \Delta T_{\rm CMB}/T_{\rm CMB}$.
Note that Eq.~(\ref{boltzeq}) has been written in the Fourier space with $k$ being the magnitude of the wave vector of the Fourier mode. The scalar variables $\delta\epsilon_b, \, \delta \epsilon_c, \, \delta\epsilon_\gamma, \, \delta u_{bS}, \, \delta u_{cS}, \, \delta u_{\gamma S}$ and $\Pi$ have been replaced by a set of variables more convenient to use, 
\begin{align}
& \delta_b = \frac{\delta \epsilon_b}{\epsilon_b}, 
\nonumber \\
& \delta_c = \frac{\delta \epsilon_c}{\epsilon_c}, 
\nonumber \\
& \Theta_0 = \frac{1}{4} \delta_\gamma = \frac{1}{4} \frac{\delta \epsilon_\gamma}{\epsilon_\gamma},
\nonumber \\
& v_b = \frac{k}{a} \delta u_{bS} ,
\nonumber \\
& v_c = \frac{k}{a} \delta u_{cS} ,
\nonumber \\
& \Theta_1 = -\frac{k}{3a} \delta u_{\gamma S}, 
\nonumber \\
& \Theta_2 = \frac{3k^2}{8a^2 \epsilon_\gamma} \Pi.
\end{align}
The interaction rate between electrons and photons is
\begin{align}
\Gamma = a n_e \sigma_T = a X_e n_b \sigma_T,
\label{interactionrate}
\end{align}
where $n_e$ and $n_b$ are the number densities of electrons and baryons, and $\sigma_T$ is the Thomson scattering cross section. The free electron fraction $X_e$ is calculated using the Saha equation before recombination and the Peebles equation during and after recombination \cite{Peebles:1968ja}.

\section{Approximate solutions at $\lowercase{a}\rightarrow 0$}
\label{app3}
For any given value of $k$, we want to solve Eq.~(\ref{boltzeq}) together with the Fourier transform of Eq.~(\ref{fieldeqs1}) numerically. The adiabatic initial condition, which is inherited from inflation, sets
\begin{align}
\delta_b \approx \delta_c \approx 3\Theta_0, 
\label{adbcon} 
\end{align}  
at $a\rightarrow 0$. To start numerical integrations at $a\rightarrow 0$, we still need initial conditions for other variables $\Psi, \, \Phi, \, \delta b_\eta, \, \delta b_S, \, v_b, \, v_c, \Theta_1, \, \Theta_2, \, ...$, namely that we need to find approximate solutions for those variables at $a\rightarrow 0$. To do this, we first simplify the equations by removing the velocity variables as well as the higher-order multipoles $\Theta_l$. Thus, Eq.~(\ref{boltzeq}) gives
\begin{align}
& \delta_b \approx 3 \Phi + A_b, 
\nonumber \\
& \delta_c \approx 3 \Phi + A_c, 
\nonumber \\
& \Theta_0 \approx \Phi + A_0, 
\label{appsol1}
\end{align}
where the integral constants satisfy $A_b=A_c=3A_0$ to be consistent with Eq.~(\ref{adbcon}), while Eq.~(\ref{fieldeqs1}) becomes a closed set of equations for $\Psi, \, \Phi, \, \delta b_\eta$ and $\delta b_S$. Once approximate solutions for $\Psi, \, \Phi, \, \delta b_\eta$ and $\delta b_S$ are obtained, they can be used in Eq.~(\ref{boltzeq}) to find approximate solutions for the velocity variables as well as the higher-order multipoles. Then, we will be able to check if the velocity variables and the higher-order multipoles contribute relevant terms in solving $\delta_b, \, \delta_c, \Theta_0$ and $\Psi, \, \Phi, \, \delta b_\eta, \, \delta b_S$ approximately. If they do contribute relevantly and cannot be neglected in the first place, we can correct the obtained approximate solutions accordingly. 

With the scheme stated, to find approximate solutions for $\Psi, \, \Phi, \, \delta b_\eta$ and $\delta b_S$ at $a\rightarrow 0$ from Eq.~(\ref{fieldeqs1}), the behaviors of ${\cal H}$ and $b_\eta$ at $a\rightarrow 0$ are necessary. Equation~(\ref{flrwsol1}) gives
\begin{align}
\cal H \approx
\begin{cases}
H_0 \sqrt{1-\Omega_{V_1}} \, a^{\alpha/2} & \text{if } \alpha < 2 ,\\ 
H_0 \sqrt{\Omega_{V_1}} \, a & \text{if } \alpha > 2 ,
\end{cases}
\label{appsol2}
\end{align}
for the spatially flat solution. Note that $\alpha=2$ corresponds to $\xi_1+4\xi_2=0$, and it turns out in this case  
\begin{align}
{\cal H} \approx H_0 \, a \sqrt{\frac{8\tilde V_1}{\xi_1} \ln{a}} \, .
\end{align}
The $\ln{a}$ factor spoils the power-law behavior of ${\cal H}$ and subsequently power-law behaviors of $\Psi, \, \Phi, \, \delta b_\eta$ and $\delta b_S$. We will avoid considering the case of $\alpha=2$. 
The behavior of $b_\eta$ at $a\rightarrow 0$ turns out to be more delicated than that of ${\cal H}$. By a careful analysis of the first equation in Eq.~(\ref{flrweq}), we find 
\begin{align}
b_\eta \approx
\begin{cases}
\sqrt{\frac{2\alpha+4}{\xi_1\left(\alpha+4\right)}} \, a  & \text{if } \alpha < -4 ,\\ 
D_1 \, a^{-\alpha/4} & \text{if } -4 < \alpha < 0 ,\\
\sqrt{\frac{\Omega_{r 0}}{\xi_2(\Omega_{V_1}-1)}} \, a^{-\alpha/2} & \text{if } 0 < \alpha < 2 , \\
\sqrt{\frac{(\alpha-2)\Omega_{r0}}{2(\alpha-3)\tilde V_1}} \, \frac{1}{a} & \text{if } 2 < \alpha < 3 , \\
D_2 \, a^{(1-\alpha)/2} & \text{if } \alpha > 3 ,
\end{cases}
\label{appsol3}
\end{align}
where $D_1$ and $D_2$ are integral constants with no presumed values. For the cases of $\alpha=-4, \, 0, \, 2$ and $3$, we find that $\ln{a}$ is encountered in the approximate solution of $b_\eta$, ruining the expected power-law behavior. We will not consider these pathological cases.

Neglecting the velocity variables and the higher-order multipoles, and using the approximate solutions in Eqs.~(\ref{appsol1}), (\ref{appsol2}) and (\ref{appsol3}), Eq.~(\ref{fieldeqs1}) in the Fourier space becomes a closed set of ODEs for $\Psi, \, \Phi, \, \delta b_\eta$ and $\delta b_S$. We expect to find approximate solutions for them also in the form of power-law relations with respect to $a$ if the time evolution of these variables is physically sensible. This turns out to be true for $\alpha<0$. For $\alpha>0$, we find that the approximate solutions for $\Psi, \, \Phi, \, \delta b_\eta$ and $\delta b_S$ consist of terms like $e^{a^r}$ with constant values of $r$, which we take as a signal indicating the cases of $\alpha>0$ being unphysical.

We find power-law approximate solutions of $\Psi, \, \Phi, \, \delta b_\eta$ and $\delta b_S$ for both the case of $\alpha<-4$ and the case of $-4<\alpha<0$. But we will focus only on the case of $-4<\alpha<0$, because the case of $\alpha<-4$ is excluded by the fact that the universe is currently expanding with an acceleration. This can be shown by using Eq.~(\ref{partrans}) to substitute $\xi_1$ in the approximate solution of $b_\eta$ for $\alpha<-4$ in Eq.~(\ref{appsol3}). The approximate solution exists when
\begin{align}
-\frac{\alpha+2q_0}{\tilde V_1 \left(\alpha+4\right)} > 0 .
\end{align}   
With $\tilde V_1>0$ and $\alpha<-4$, the inequality cannot hold if the deceleration parameter $q_0$ is negative. Numerical integration also verifies that for $\alpha<-4$ and $q_0<0$, $b_\eta$ becomes imaginary when $a$ approaches $0$.

The approximate solutions of $\Psi, \, \Phi, \, \delta b_\eta$ and $\delta b_S$ that we find for the case of $-4<\alpha<0$ take the form
\begin{align}
& \Psi \approx c_1 \, a^r ,
\nonumber \\
& \Phi \approx c_2 \, a^r ,
\nonumber \\
& \delta b_\eta \approx c_3 \, a^{r-\frac{\alpha}{4}} ,
\nonumber \\
& \delta b_S \approx c_4 \, a^{r-\frac{3\alpha}{4}} ,
\label{appsol4}
\end{align}
where $c_1, \, c_2, \, c_3, \, c_4$ and $r$ are constants. Notice that (i) $\Psi$ can be eliminated using the second equation in Eq.~(\ref{fieldeqs1}), leading to two first-order ODEs for $\Phi$ and $\delta b_\eta$ and one second-order ODE for $\delta b_S$ in the Fourier space, and (ii) the ODEs are inhomogeneous as substituting $\delta \epsilon_b, \, \delta\epsilon_c$ and $\delta\epsilon_\gamma$ using Eq.~(\ref{appsol1}) introduces source terms proportional to $A_0$. Therefore, we find 4 homogeneous solutions and one inhomogeneous solution as follows.
\begin{enumerate}
\item Homogeneous solution I: $r$ takes the value
\begin{align}
r_1 = 0 ,
\end{align}
and the coefficients are 
\begin{align}
& c_1 = -\frac{c_4 (\alpha-2) H_0 \sqrt{1-\Omega_{V_1}}}{2 D_1},
\nonumber \\
& c_2 = \frac{c_4 (\alpha+3) H_0 \sqrt{1-\Omega_{V_1}}}{D_1} ,
\nonumber \\
& c_3 = -\frac{3}{4} c_4 \alpha H_0 \sqrt{1-\Omega_{V_1}} .
\end{align}

\item Homogeneous solution II: $r$ takes the value
\begin{align}
r_2 = \frac{\alpha}{2} + \sqrt{\frac{\alpha^2}{4} + \frac{\left(2-\alpha\right)\xi_1}{\kappa}} ,
\end{align}
and the coefficients are 
\begin{align}
c_1 =&\, \frac{c_4 (\alpha+2) H_0  \sqrt{1-\Omega_{V_1}} }{2 \alpha D_1 \kappa} 
\nonumber \\
& \times \left(\sqrt{\kappa \left(\alpha^2 \kappa-16 (\alpha-2) \xi_1\right)}+4 \kappa\right),
\nonumber \\
c_2 =&\, -\frac{2 c_4 (\alpha+2) H_0 \sqrt{1-\Omega_{V_1}}}{\alpha D_1} ,
\nonumber \\
c_3 =&\, \frac{c_4 (\alpha+2) H_0 \sqrt{1-\Omega_{V_1}}  }{2 \alpha \kappa} 
\nonumber \\
& \times \left(\sqrt{\kappa \left(\alpha^2 \kappa-16 (\alpha-2) \xi_1\right)}-\alpha \kappa\right).
\end{align}

\item Homogeneous solution III: $r$ takes the value
\begin{align}
r_3 = \frac{\alpha}{2} - \sqrt{\frac{\alpha^2}{4} + \frac{\left(2-\alpha\right)\xi_1}{\kappa}} ,
\end{align}
and the coefficients are 
\begin{align}
c_1 =&\, -\frac{c_4 (\alpha+2) H_0 \sqrt{1-\Omega_{V_1}} }{2 \alpha D_1 \kappa}
\nonumber \\
& \times \left(\sqrt{\kappa \left(\alpha^2 \kappa-16 (\alpha-2) \xi_1\right)}-4 \kappa\right) ,
\nonumber \\
c_2 =&\, -\frac{2 c_4 (\alpha+2) H_0 \sqrt{1-\Omega_{V_1}}}{\alpha D_1} ,
\nonumber \\
c_3 =&\, -\frac{c_4 (\alpha+2) H_0 \sqrt{1-\Omega_{V_1}}}{2 \alpha \kappa}
\nonumber \\
& \times \left(\sqrt{\kappa \left(\alpha^2 \kappa-16 (\alpha-2) \xi_1\right)}+\alpha \kappa\right) .
\end{align}

\item Homogeneous solution IV: $r$ takes the value
\begin{align}
r_4 = \alpha ,
\end{align}
and the coefficients are 
\begin{align}
& c_1 = \frac{c_4  (\alpha+2) H_0 \sqrt{1-\Omega_{V_1}}}{2 D_1}  ,
\nonumber \\
& c_2 = -\frac{c_4  H_0 \sqrt{1-\Omega_{V_1}}}{D_1},
\nonumber \\
& c_3 = \frac{1}{4} c_4 \alpha H_0 \sqrt{1-\Omega_{V_1}}  .
\end{align}

\item Inhomogeneous solution: $r$ takes the value
\begin{align}
r_0 = -\frac{\alpha}{2} ,
\end{align}
and the coefficients are 
\begin{align}
c_1 =&\, -\frac{16 (\alpha-2) (\alpha-1) (\alpha+2) \Omega_{r0}A_0}{3 \alpha (\alpha+4) D_1^2 \left(15 \alpha^2 \kappa+16 (\alpha-2) \xi_1\right)}
\nonumber \\
& \times \frac{3 \alpha (\alpha+12) \kappa-16 (\alpha-2) \xi_1}{ 2 (\alpha+2) \tilde V_1+(\alpha-2) \xi_1} ,
\nonumber \\
c_2 =&\, -\frac{16 (\alpha-2)(\alpha+2) \Omega_{r0}A_0}{3 \alpha (\alpha+4) D_1^2 \left(15 \alpha^2 \kappa+16 (\alpha-2) \xi_1\right) }
\nonumber \\
& \times \frac{3 \alpha (\alpha+12) \kappa-16 (\alpha-2) \xi_1}{ 2 (\alpha+2) \tilde V_1+(\alpha-2) \xi_1 } ,
\nonumber \\
c_3 =&\, \frac{8 (\alpha-2)(\alpha+2)  \Omega_{r0}A_0}{3 \alpha (\alpha+4) D_1 \left(15 \alpha^2 \kappa+16 (\alpha-2) \xi_1\right) }
\nonumber \\
& \times \frac{ 15 \alpha^3 \kappa+32 (\alpha-2) (2 \alpha+3) \xi_1 }{ 2 (\alpha+2) \tilde V_1+(\alpha-2) \xi_1 },
\nonumber \\
c_4 =&\, -\frac{32 (\alpha+2) \Omega_{r0}A_0}{3 \alpha (\alpha+4) D_1 \left(15 \alpha^2 \kappa+16 (\alpha-2) \xi_1\right) }
\nonumber \\
& \times \frac{ 3 \alpha^2 \kappa+8 (\alpha-2) \xi_1 }{ H_0 \xi_1\left(1-\Omega_{V_1}\right)^{3/2} } .
\label{inhomoappsol}
\end{align}

\end{enumerate}

With the approximate solutions of $\Psi$ and $\Phi$, approximate solutions of the velocity variables and the higher-order multipoles can be found from Eq.~(\ref{boltzeq}). First, the equation for $v_c$ can be directly solved to get 
\begin{align}
v_c \approx - \frac{k c_1}{H_0 \left(1-\frac{\alpha}{2}+r \right) \sqrt{1-\Omega_{V_1}}} a^{-\frac{\alpha}{2}+r} .
\end{align}
Then, for $v_b$ and $\Theta_1$, using the tight-coupling approximation
\begin{align}
v_b \approx -3\Theta_1,
\end{align}
the equation for $\Theta_1$ becomes
\begin{align}
\Theta_1' \approx \frac{k}{3} \left( \Phi+\Psi \right) + \frac{kA_0}{3} .
\end{align} 
For the homogeneous solutions of $\Phi$ and $\Psi$, we find
\begin{align}
\Theta_1 \approx \frac{k(c_1+c_2)}{3H_0 \left( r-\frac{\alpha}{2} \right) \sqrt{1-\Omega_{V_1}}} a^{-\frac{\alpha}{2}+r} ,
\end{align}
and for the inhomogeneous solution of $\Phi$ and $\Psi$, we find
\begin{align}
\Theta_1 \approx -\frac{2 k A_0}{3\alpha H_0  \sqrt{1-\Omega_{V_1}}} a^{-\frac{\alpha}{2}}.
\label{appsol6}
\end{align}
Finally, the equations for the higher-order multipoles give the recurrence relations 
\begin{align}
& \Theta_2 \approx \frac{4k}{9} \frac{a^2}{\Gamma_0} \Theta_1,
\nonumber \\
& \Theta_l \approx \frac{lk}{2l+1} \frac{a^2}{\Gamma_0} \Theta_{l-1}, \quad l\ge 3 ,
\label{appsol5}
\end{align}
where we have used $\Gamma \approx \Gamma_0/a^2$ at $a\rightarrow 0$, with $\Gamma_0$ being a constant.

Having the approximate solutions for the velocity variables and the higher-order multipoles, we need to check if neglecting them at the beginning when solving $\delta_b, \, \delta_c, \Theta_0$ and $\Psi, \, \Phi, \, \delta b_\eta, \, \delta b_S$ is justified or not. It is straightforward to verify that the velocity variables and the higher-order multipoles can be neglected in solving the approximate solutions of $\delta_b, \, \delta_c, \Theta_0$ and the homogeneous solutions of $\Psi, \, \Phi, \, \delta b_\eta, \, \delta b_S$, while the photon velocity variable $\delta u_{\gamma S} = -3a \Theta_1/k$ does contribute relevant terms in solving the inhomogeneous solutions of $\Psi, \, \Phi, \, \delta b_\eta, \, \delta b_S$. Knowing that the inhomogeneous solutions of $\Psi, \, \Phi, \, \delta b_\eta, \, \delta b_S$ take the form of Eq.~(\ref{appsol4}) with $r=r_0=-\alpha/2$, the coefficients $c_1, \, c_2, \, c_3, \, c_4$ can be recalculated taking into consideration of the contribution of $\delta u_{\gamma S}$. In fact, we have shown in Eq.~(\ref{inhomoappsol}) the correct coefficients for the inhomogeneous solutions of $\Psi, \, \Phi, \, \delta b_\eta, \, \delta b_S$, as it is unnecessary to show the inaccurate result obtained with $\delta u_{\gamma S}$ dropped.

Equipped with the above approximate solutions at $a\rightarrow 0$, numerical integrations for solving the ODE system consists of Eq.~(\ref{boltzeq}) and Eq.~(\ref{fieldeqs1}) in the Fourier space can be carried out from a small enough value of $a$ ($\ln{a}=-14$ in our numerical code), with the initial values of the variables given by the sum of the 5 approximate solutions. In fact, we need to point out that out of the 5 approximate solutions, only the inhomogeneous solution is relevant for the CMB calculation, because it is the mode corresponding to an initial perturbation in the CMB photons' energy density. The 4 homogeneous modes, including one mode corresponding to an initial metric perturbation and three modes corresponding to initial perturbations in the bumblebee field, turn out to contribute insignificantly in calculating the CMB anisotropy. This is similar to GR where there is no mode of bumblebee perturbation and therefore only one homogeneous mode corresponding to an initial metric perturbation.    

The perturbation equations in Eq.~(\ref{fieldeqs1}) apply to GR by setting $b_\eta = \delta b_\eta = \delta b_S= 0$ and adding $\epsilon_\Lambda = \Lambda/\kappa$ to the combination $\epsilon_b+\epsilon_c+\epsilon_\gamma$ in the third equation. Using the above discussed method, one homogeneous solution can be found by closing the equation of $\Phi$ using $\Theta_0 \approx \Phi$,
\begin{align}
& \Psi \approx \Phi \approx \Theta_0 \approx \frac{c_1}{a^3} ,
\nonumber \\
& \Theta_1 \approx -\frac{kc_1}{3H_0 \sqrt{\Omega_{r0}}} \frac{1}{a^2}, 
\nonumber \\
& v_b \approx v_c \approx \frac{kc_1}{ H_0 \sqrt{\Omega_{r0}} } \frac{1}{a^2} ,
\label{grappsol1}
\end{align}
where $c_1$ is an arbitrary constant, and one inhomogeneous solution can be found by closing the equation of $\Phi$ using $\Theta_0 \approx \Phi+A_0$,  
\begin{align}
& \Psi \approx \Phi \approx -\frac{2A_0}{3} , 
\nonumber \\
& \Theta_0 \approx \frac{A_0}{3} ,
\nonumber \\
& \Theta_1 \approx -\frac{kA_0}{9H_0 \sqrt{\Omega_{r0}}} a ,
\nonumber \\
& v_b \approx v_c \approx \frac{kA_0}{3H_0 \sqrt{\Omega_{r0}}} a  .
\label{grappsol2}
\end{align}
For both solutions, the higher-order multipoles are given by Eq.~(\ref{appsol5}). A set of general initial data is a combination of the homogeneous solution in Eq.~(\ref{grappsol1}) and the inhomogeneous solution in Eq.~(\ref{grappsol2}). But as the homogeneous solution contributes insignificantly, it is sufficient to use only the inhomogeneous solution to set up the initial condition for numerically integrating the full set of perturbation equations.

\section{Checking our CMB code by applying it to the $\Lambda$CDM model}
\label{app4}
Using the conformal expansion rate for the $\Lambda$CDM model in GR as shown in Eq.~(\ref{flrwlcdm}), 
switching off $b_\eta, \, \delta b_\eta, \, \delta b_S$, and adding $\epsilon_\Lambda=\Lambda/\kappa$ in Eq.~(\ref{fieldeqs1}), our CMB code applies to the $\Lambda$CDM model with $\Omega_{{\cal K}0}=0$. By calculating the CMB power spectrum in the $\Lambda$CDM model and comparing with results from CAMB~\footnote{\url{https://camb.readthedocs.io/en/latest/}}, we can have an estimate for the error in the results produced using our code. 

\begin{figure*}
 \includegraphics[width=0.9\linewidth]{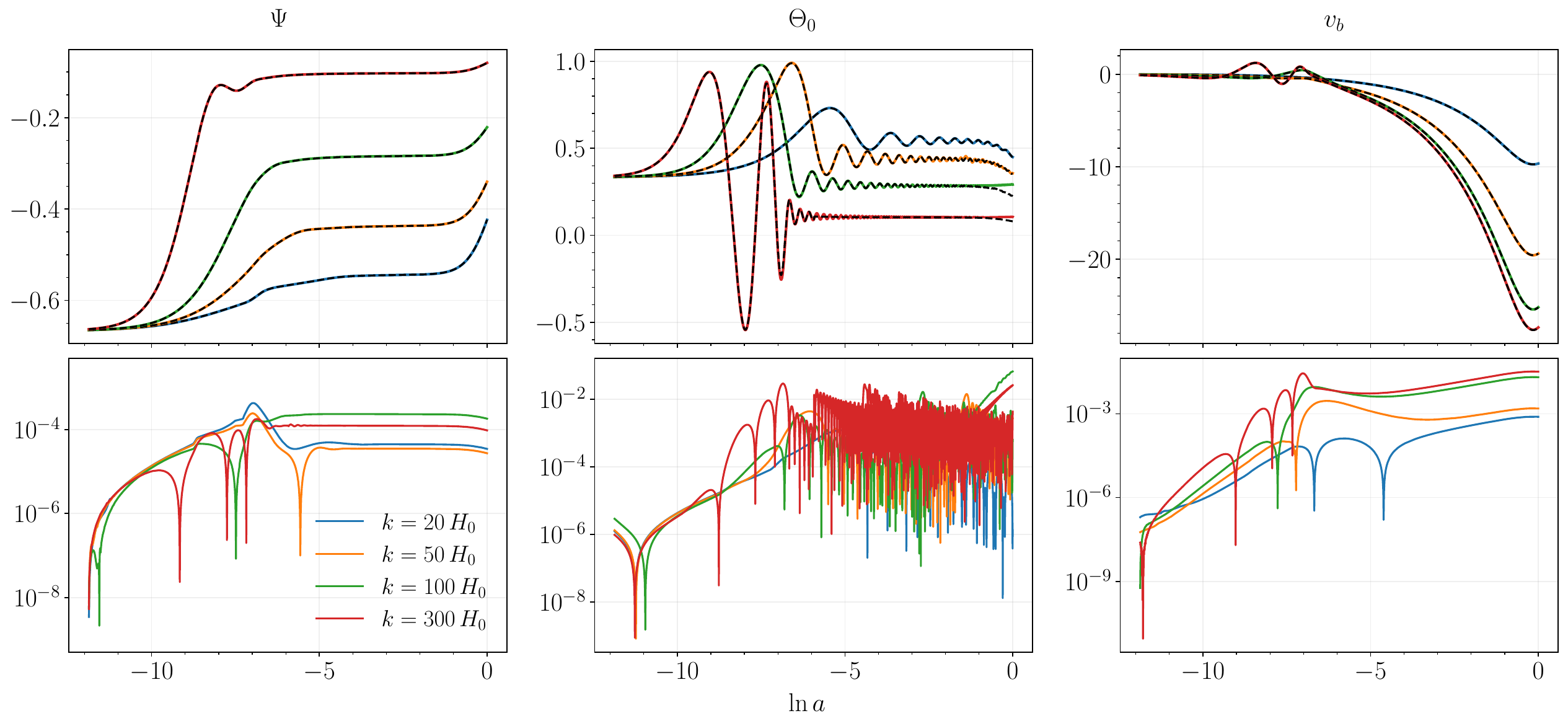}
 \caption{Examples of the numerical solutions for the representative perturbation variables $\Psi, \, \Theta_0$ and $v_b$. Upper row: the solutions. Lower row: their absolute errors checked against CAMB. Four Fourier modes with different values of $k$ are shown. The colored lines in the upper row are results calculated using our code, while the black dashed lines are results from CAMB.     }
\label{app_fig1}
\end{figure*}

\begin{figure*}
 \includegraphics[width=0.9\linewidth]{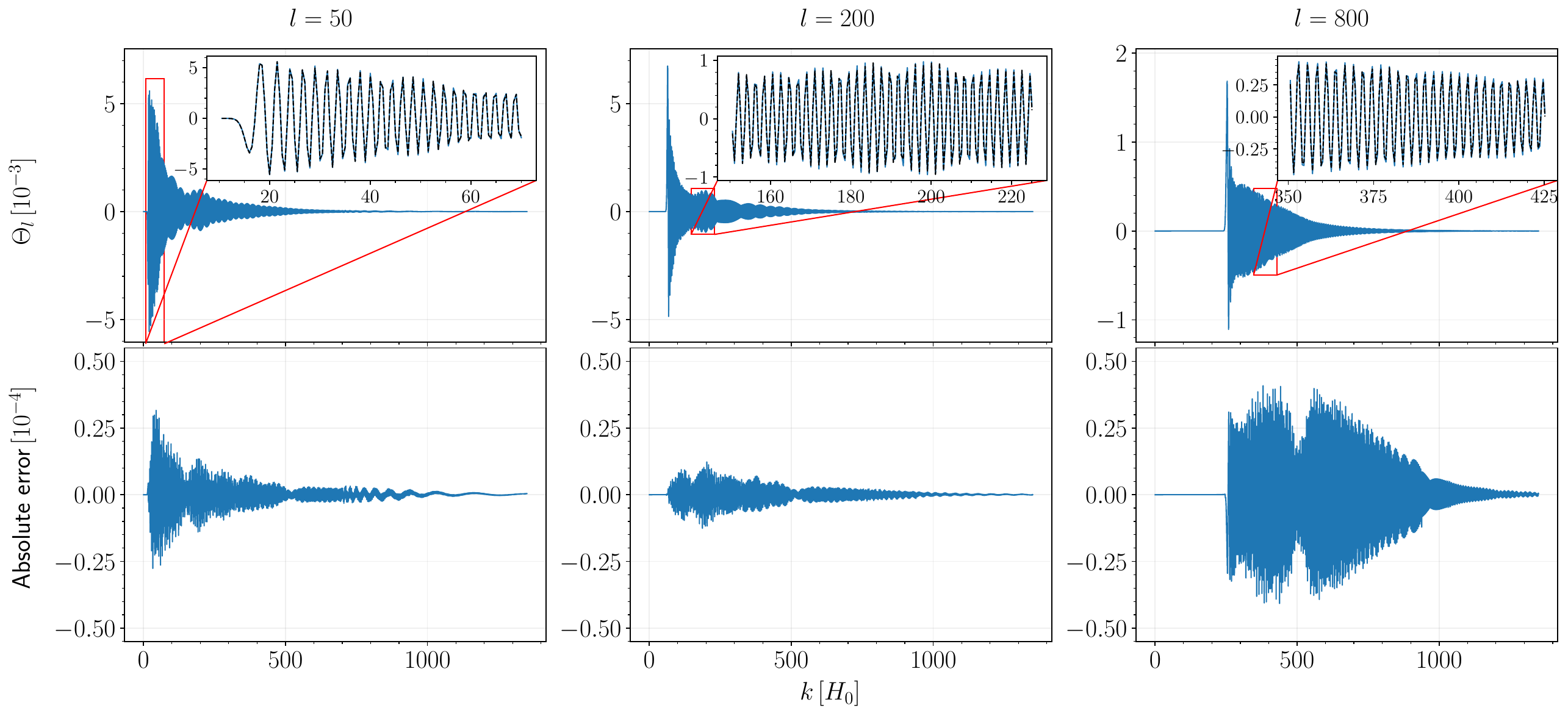}
 \caption{Upper row: examples of the transfer function $\Theta_l$. Lower row: their absolute errors checked against CAMB. The insets in the upper row zoom in on the part of $\Theta_l$ where the errors are the largest. In the insets, the black dashed lines are results from CAMB, while the solid lines are our results.    }
\label{app_fig3}
\end{figure*}

\begin{figure}
 \includegraphics[width=0.9\linewidth]{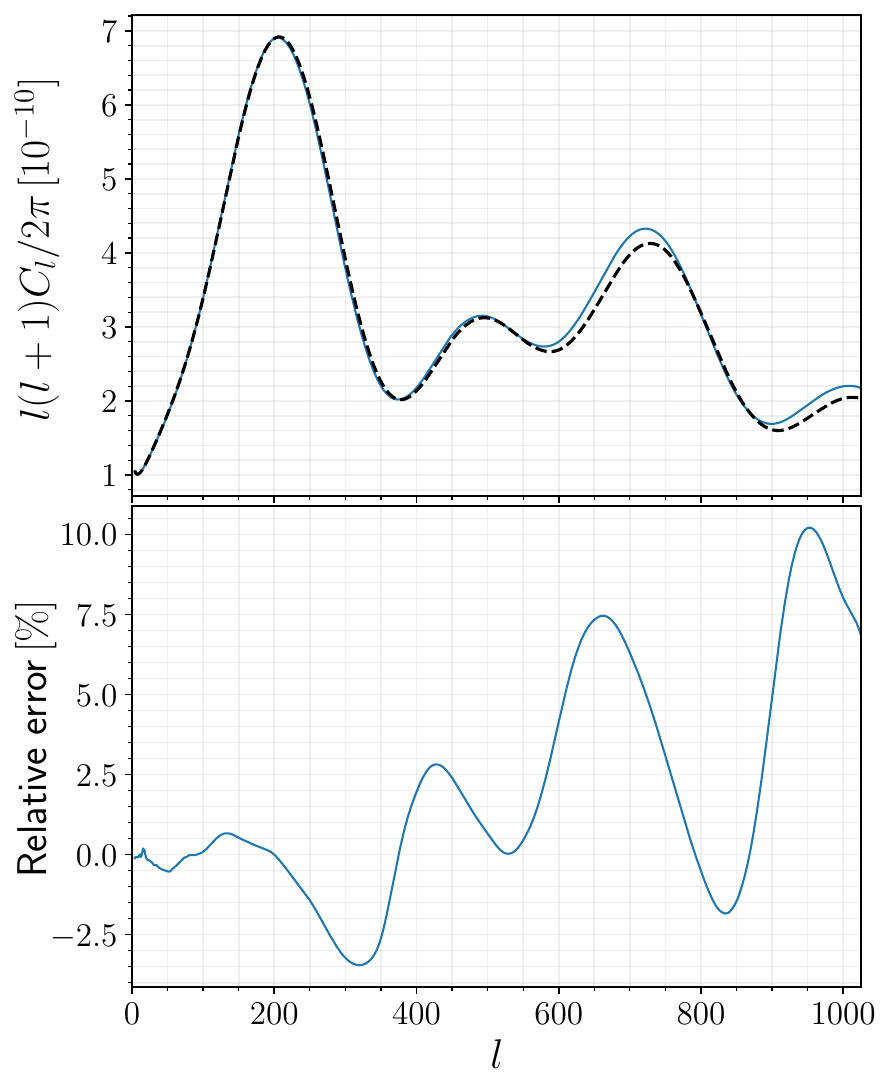}
 \caption{Upper panel: An example of the CMB temperature power spectrum $C_l$ calculated using our code (the solid line) and the corresponding result from CAMB (the dashed line). Lower panel: the relative error of our result checked against the result from CAMB.   }
\label{app_fig5}
\end{figure}

As explained in Section~\ref{sec:constraints}, there are three steps in calculating the CMB temperature power spectrum and the results can be shown at three levels: (i) the solutions for the perturbation variables, (ii) the transfer function $\Theta_l(k)$, and (iii) the final power spectrum $C_l$.
Figures~\ref{app_fig1}-\ref{app_fig5} show comparisons between results from our code and from CAMB. The parameters of the $\Lambda$CDM model calculated are
\begin{align}
& H_0 = 70 \, {\rm km/s/Mpc}, 
\nonumber \\
& \Omega_{b0}=0.05, 
\nonumber \\
& \Omega_{c0}=0.25, 
\nonumber \\
& T_{\rm CMB}=2.7\, {\rm K} ,
\nonumber \\
& n_s = 1 .
\end{align} 
Neutrinos and reionization are switched off when using CAMB to match the simple cosmological model in our code.

\section*{References}

\bibliography{mybibfile}

\end{document}